\documentclass[11pt,a4paper]{article}

\usepackage[left=2.35cm,top=3cm,bottom=3cm,right=2.3cm]{geometry}

\usepackage{graphicx}
\usepackage[svgnames,x11names]{xcolor}
\definecolor{lightgray}{HTML}{D5D5D5}
\definecolor{lightergray}{HTML}{EEEEEE}
\usepackage{upgreek}
\usepackage{cite}
\usepackage{xcolor,colortbl}
\usepackage{slashed}

\definecolor{col}{rgb}{.4,.4,1}
\definecolor{col1}{rgb}{.4,.4,1}
\definecolor{col2}{rgb}{.4,.5,1}
\definecolor{col3}{rgb}{.4,.6,1}
\definecolor{col4}{rgb}{.4,.7,1}
\definecolor{hard}{rgb}{0.56,0.69,0.19}
\definecolor{stuff}{rgb}{0.88,0.88,0.88}
\definecolor{soft}{rgb}{0.88,0.61,0.14}
\definecolor{math1}{rgb}{0.37,0.51,0.71}
\definecolor{math2}{rgb}{0.88,0.61,0.14}
\definecolor{math3}{rgb}{0.56,0.69,0.19}
\definecolor{math4}{rgb}{0.92,0.39,0.21}
\definecolor{math5}{rgb}{0.53,0.47,0.70}
\definecolor{charge}{rgb}{0.8 0.15 0.15}

\usepackage{amsmath}
\usepackage{mathtools}
\usepackage{amsfonts}
\usepackage{amssymb}
\usepackage{slashed}
\usepackage{enumerate}
\usepackage[caption=false]{subfig}
\usepackage[colorlinks=true
,urlcolor=blue
,anchorcolor=blue
,citecolor=blue
,filecolor=blue
,linkcolor=blue
,menucolor=blue
,linktocpage=true
,pdfproducer=medialab
]{hyperref}
\numberwithin{equation}{section}

\usepackage{listings}
\usepackage{pgf}

\newcommand{\pder}[2][]{\frac{\partial#1}{\partial#2}}

\usepackage{xspace}
\def\CDR{\text{\scshape cdr}\xspace}
\def\HV{\text{\scshape hv}\xspace}
\def\FDH{\text{\scshape fdh}\xspace}

\usepackage{tikz}
\usetikzlibrary{decorations.pathmorphing,positioning,decorations.pathreplacing,shapes,calc}
\usetikzlibrary{decorations.pathreplacing}
\usetikzlibrary{decorations.markings}
\usetikzlibrary{arrows}
\tikzset{
    photon/.style={decorate, decoration={snake,amplitude=1.5pt,segment length=6pt}},
    tightphoton/.style={decorate, decoration={snake,amplitude=1.5pt,segment length=5pt}},
    zigzag it/.style={decorate, decoration=zigzag},
    gluon/.style={decorate, draw=black,decoration={coil,amplitude=4pt, segment length=5pt}},
    tightgluon/.style={decorate, draw=black,decoration={coil,amplitude=2pt, segment length=3pt}}
}
\def\centerarc[#1](#2,#3)(#4:#5:#6)% Syntax: [draw options] (center) (initial angle:final angle:radius)
    { \draw[#1] ({#2+#6*cos(#4)},{#3+#6*sin(#4)}) arc (#4:#5:#6); }

\newcommand\eik{\mathcal{E}}

\usepackage{amsmath,graphicx}

\newcommand{\wideeq}[2][1.5]{
  \mathrel{\overset{#2}{\scalebox{#1}[1]{$=$}}}
}
\newcommand*\pFqskip{8mu}
\catcode`,\active
\newcommand*\pFq{\begingroup
        \catcode`\,\active
        \def ,{\mskip\pFqskip\relax}%
        \dopFq
}
\catcode`\,12
\def\dopFq#1#2#3#4#5{%
        {}_{#1}F_{#2}\biggl[\genfrac..{0pt}{}{#3}{#4};#5\biggr]%
        \endgroup
}

\begin{document}
\thispagestyle{empty}

\begin{flushright}
PSI-PR-21-29\\
ZU-TH 59/21\\
IPPP/21/55
\end{flushright}
\vspace{3em}
\begin{center}
{\Large\bf Universal structure of radiative QED amplitudes at one loop}
\\
\vspace{3em}
{\sc
T.\,Engel$^{a,b}$,
A.\,Signer$^{a,b}$,
Y.\,Ulrich$^{c}$
}\\[2em]
{\sl ${}^a$ Paul Scherrer Institut,\\
CH-5232 Villigen PSI, Switzerland \\
\vspace{0.3cm}
${}^b$ Physik-Institut, Universit\"at Z\"urich, \\
Winterthurerstrasse 190,
CH-8057 Z\"urich, Switzerland \\
\vspace{0.3cm}
${}^c$ Institute for Particle Physics Phenomenology, University of
Durham, \\
South Road, Durham DH1 3LE, United Kingdom}
\setcounter{footnote}{0}
\end{center}
\vspace{6ex}

\begin{center}
\begin{minipage}{15.3truecm}
{
We present two novel results about the universal structure of
radiative QED amplitudes in the soft and in the collinear limit.  On
the one hand, we extend the well-known Low-Burnett-Kroll theorem to
the one-loop level and give the explicit relation between the
radiative and  non-radiative amplitude at subleading power in the soft
limit. On the other hand, we consider a factorisation formula at
leading power in the limit where the emitted photon becomes collinear
to a light fermion and provide the corresponding one-loop splitting
function.  In addition to being interesting in their own right these
findings are particularly relevant in the context of
fully-differential higher-order QED calculations. One of the main
challenges in this regard is the numerical stability of radiative
contributions in the soft and collinear regions. The results presented
here allow for a stabilisation of real-virtual amplitudes in these
delicate phase-space regions by switching to the corresponding
approximation without the need of explicit computations.  }
\end{minipage}
\end{center}

\newpage

\section {Introduction}

The limit of scattering amplitudes where an external massless gauge
boson becomes soft or collinear to a fermion has important
applications from a conceptual quantum field theory point of view, but
also for concrete applications in higher-order calculations. In this
paper we consider QED with massive fermions and using a diagrammatic
approach provide new explicit results of the limiting behaviour of
one-loop radiative amplitudes with practical applications in mind.

In order to meet the experimental precision the calculation of
next-to-leading order (NLO) or even next-to-next-to-leading order
(NNLO) QED corrections has become mandatory. A lot of work has
therefore been put into the computation of the corresponding loop
amplitudes where the presence of non-vanishing fermion masses results
in significant complexity. Furthermore, the need to account for
non-trivial detector geometries and acceptances has triggered the
development of numerous Monte-Carlo codes that allow for the
calculation of fully differential observables.

Most of these calculations face the difficulty of a strong scale
hierarchy due to small fermion masses acting as regulators of
collinear singularities. It is therefore one of the main challenges of
fully differential higher-order QED calculations to achieve
numerically stable implementations of the amplitudes. Particularly
delicate in this regard are radiative loop amplitudes evaluated in the
phase-space region where the emitted photon is soft. These numerical
instabilities are further exacerbated if the photon becomes collinear
to a light on-shell fermion. The approximation of the corresponding
expression using the soft expansion up to subleading power yields an
elegant solution to this problem. This next-to-soft stabilisation has
facilitated the first calculation of fully differential NNLO QED
corrections to Bhabha~\cite{Banerjee:2021mty} and
M{\o}ller~\cite{Banerjee:2021qvi} scattering. Even though the soft
expansion can straightforwardly be computed with the method of
regions~\cite{Beneke:1997zp} the corresponding calculation is
cumbersome. The currently limited knowledge of universal soft and
collinear behaviour in QED makes therefore a systematic study highly
desirable. Hence, in this article we present two universal results for radiative amplitudes at one loop, the next-to-soft limit and the collinear limit.

It is a well-known fact that in QED radiative amplitudes exhibit
universal factorisation at leading power in the soft limit given by
the eikonal approximation to any order in perturbation
theory~\cite{Yennie:1961ad}. Furthermore, it has been shown a long
time ago by Low, Burnett, and Kroll~\cite{Low:1958sn,Burnett:1967km}
that also the subleading term is related to the non-radiative
amplitude at tree level via a differential operator.\footnote{In the
case of gravity this even holds true up to sub-subleading
power~\cite{Cachazo:2014fwa,Bern:2014vva,Beneke:2021umj}.} This
so-called LBK theorem was later extended to massless
particles~\cite{DelDuca:1990gz} where a universal radiative jet
function was introduced to take into account collinear effects. More
recently, the massless version of the theorem has attracted some
attention in the context of resummation of next-to-leading power
threshold logarithms. To this end the theorem was extended to also
include loop corrections in the framework of diagrammatic
factorisation~\cite{Bonocore:2015esa,Bonocore:2016awd,Laenen:2020nrt}
as well as in soft-collinear effective
theory (SCET)~\cite{Larkoski:2014bxa,Beneke:2019oqx,Liu:2021mac}. In this paper, however, we are interested in the case of QED where all fermion masses and all other scales are considered to be much larger than the energy of the emitted photon. In this case, these recent loop-level extensions are not applicable since the underlying effective theory is heavy-quark effective theory (HQET) and not SCET. In particular, there is no radiative jet function in this scenario due to the absence of any collinear scale. This leaves hard and soft modes as the only relevant degrees of freedom.

The collinear limit of radiative amplitudes has been extensively
investigated in the context of QCD with massless quarks where it gives
rise to infrared singularities. The factorisation into a process-independent
splitting function multiplying the non-radiative amplitude
has therefore been known for some time now~\cite{Bern:1994zx,
Kosower:1999rx}. While the splitting functions correspond to the
Altarelli-Parisi kernels at tree level, this is no longer true if loop
corrections are taken into account. The two-loop corrections to the
QCD splitting functions have been calculated
in~\cite{Bern:2004cz,Badger:2004uk}. Unfortunately, much less is known in the
case of QED where collinear divergences are regularised by finite
fermion masses. Taking these masses to be small, we expect a similar factorising structure. This is due to the applicability of SCET to the case of small but non-vanishing fermion masses. Nevertheless, the corresponding splitting function is currently only known at tree level where it coincides with the QCD version up to an additional mass term~\cite{Baier:1973ms,Berends:1981uq,Kleiss:1986ct,Dittmaier:1999mb}.

It is therefore the goal of this paper to present the one-loop
generalisation of both, the LBK theorem as well as the small-mass
collinear factorisation formula in QED. The paper is organised as
follows: We start by giving the most important conventions and
definitions in Section~\ref{sec:notation}. Section~\ref{sec:soft} then
discusses the soft limit at subleading power while we present our
findings about the leading power collinear limit in
Section~\ref{sec:collinear}. In both cases we first give a brief
summary of the tree-level derivations before presenting the one-loop
extension. The main results of this paper are given
in~\eqref{eq:lbk_oneloop},~\eqref{eq:collfac_isr}
and~\eqref{eq:collfac_fsr}. In Section~\ref{sec:validation} we then
apply and validate these formulas in the process $e^{-}e^{+}\to
e^{-}e^{+}\gamma\gamma$. Finally, we comment on possible applications
and future developments in Section~\ref{sec:conclusion}.

\section{Notation and conventions}
\label{sec:notation} 

We denote the amplitude for a process with $n$ final state particles
by $\mathcal{A}_n$ and the corresponding QED $l$-loop correction by
$\mathcal{A}_n^{(l)}$. Analogously, we use $\mathcal{M}_n$ and
$\mathcal{M}_n^{(l)}$ for the unpolarised squared amplitude and refer
to it as matrix element. The correponding quantities for the radiative
process with one additional photon in the final state are given by
$n\to n+1$. The perturbative expansion of the amplitudes and matrix
elements is done in terms of the fermion charge $Q$ where $Q=-e$ for
an incoming particle or an outgoing antiparticle and $Q=+e$ otherwise.
The symbol $\Gamma$ is used in various places to generically
parametrise part of an amplitude.

In what follows we always denote the loop momentum by $\ell$. Furthermore, we consistently take $k$ to represent the momentum of the emitted photon in a radiative process. It is the soft and
collinear limit of this particle that is studied in the following
sections. Other on-shell momenta are denoted by
$p_i$, i.e. $p_i^2=m_i^2$ with $m_i$ the particle mass. The
corresponding velocity is then given by
$\beta_i=(1-m_i^2/E_i^2)^{1/2}$ where $E_i$ is the energy. In the situation where a particle may also be off shell we
use $q_i$ instead.

To make the power counting in the soft expansion transparent we
introduce the soft book-keeping parameter $\xi$. The soft limit is
then governed by the scaling $k \sim \xi \ll m_i \sim \xi^0$ which yields the expansion
\begin{align}
    \mathcal{M}_{n+1}
    \wideeq{k \sim \xi}
    \frac{1}{\xi^2}\mathcal{M}_{n+1}^\text{sLP} 
    + \frac{1}{\xi}\mathcal{M}_{n+1}^\text{sNLP}
    + \mathcal{O}(\xi^0),
\end{align}
with the leading and subleading power contributions
$\mathcal{M}_{n+1}^\text{sLP}$ and $\mathcal{M}_{n+1}^\text{sNLP}$.

Similarly, we introduce the book-keeping parameter $\lambda$ for the
power counting in the collinear limit. The corresponding expansion
is only meaningful if the collinear fermion is light. We therefore
take the mass to scale as $p^2 = m^2 \sim \lambda^2$.
This in turn gives the behaviour $k \cdot p \sim \lambda^2$ due to
\begin{align}\label{eq:kp_scaling}
    k\cdot p 
    \wideeq{\sphericalangle(k, p)\to 0} E_k E_p (1-\beta_p)
    \wideeq{m^2\sim\lambda^2} 
    E_k E_p \Big(\lambda^2 \frac{1}{2}\frac{m^2}{E_p^2} + \mathcal{O}(\lambda^4) \Big) .
\end{align}
At the level of the matrix element this gives rise to the expansion
\begin{align}
    \mathcal{M}_{n+1}
    \wideeq{k\cdot p \sim \lambda^2}
    \frac{1}{\lambda^2}\mathcal{M}_{n+1}^\text{cLP} 
    + \mathcal{O}(\lambda^{-1}).
\end{align}
In the collinear limit the amplitudes scale as
$\mathcal{A}_{n+1} \sim \lambda^{-1}$ and not
as $\sim \lambda^{-2}$ as one would naively expect.

The parameters $\xi$ and $\lambda$ are just used to facilitate the
book keeping of the corresponding expansions. In all our equations
their numerical value is $\xi=\lambda=1$. Furthermore, we emphasize that the different treatment of the fermion masses in the two limits has crucial implications. The small mass in the collinear limit introduces a collinear scale that is absent in the soft limit. From a more formal effective field theory perspective, the collinear case is therefore governed by SCET while the soft behaviour can be described in terms of HQET.

It is well known~\cite{Yennie:1961ad} that the leading-power term in the soft
expansion is given by
\begin{equation}\label{eq:eikonalapprox}
    \mathcal{M}_{n+1}^\text{sLP} = \eik \mathcal{M}_{n}
\end{equation}
to all orders in perturbation theory with the eikonal factor
\begin{equation}
    \eik = -\sum_{ij} Q_i Q_j \frac{p_i\cdot p_j}{(k\cdot p_i)(k\cdot p_j)}
\end{equation}
summing over all external legs. It is the purpose of this paper to
present similar universal results for the subleading term
$\mathcal{M}_{n+1}^\text{sNLP}$ (Section~\ref{sec:soft}) as well as
the leading collinear contribution $\mathcal{M}_{n+1}^\text{cLP}$
(Section~\ref{sec:collinear}).

\section{Soft factorisation at subleading power}
\label{sec:soft} 

It has been shown a long time ago by Low, Burnett, and Kroll that the
soft expansion of radiative tree-level amplitudes can be related up to
subleading power to the non-radiative process by means of a
differential operator~\cite{Low:1958sn,Burnett:1967km}. In this
section we will generalise this LBK theorem to the one-loop level. We
first start in Section~\ref{sec:lbk} by giving a short review of the
tree-level derivation. The one-loop extension is then discussed in
detail in the following Section~\ref{sec:lbk_oneloop} with the main
result given in~\eqref{eq:lbk_oneloop}.

\subsection{The LBK theorem}
\label{sec:lbk}

Following~\cite{Burnett:1967km} we split the radiative tree-level
amplitude $\mathcal{A}^{(0)}_{n+1}$ into contributions due to external
and internal emission
\begin{equation}
    \mathcal{A}_{n+1}^{(0)} =
    \sum_i \Bigg(
    \begin{tikzpicture}[scale=.7,baseline={(1,0)}]
    	
    \draw[line width=.3mm] (-1,-1)-- (0,0) -- (1,-1);
    \draw[line width=.3mm]  (-1,+1) node[right]{} -- (0,0) -- (1,+1);
    \draw[line width=.3mm]  [tightphoton] (0,+1) node[right]{$k$} -- (-.7,.7);
    
    \draw[line width=.3mm]  [fill=stuff] (0,0) circle (0.6) node[]{$\Gamma^\text{ext}$};
    \node at (-1.2,.7) {$p_i$};
    \end{tikzpicture}
    \hspace{0.185cm} \Bigg)
    +
    \begin{tikzpicture}[scale=.7,baseline={(1,0)}]
    	
    \draw[line width=.3mm] (-1,-1)-- (0,0) -- (1,-1);
    \draw[line width=.3mm]  (-1,+1) -- (0,0) -- (1,+1);
    \draw[line width=.3mm]  [tightphoton] (0,+1.2) node[right]{$k$} -- (0,0);
    
    \draw[line width=.3mm]  [fill=stuff] (0,0) circle (0.6);
    \end{tikzpicture}
    =
    \mathcal{A}_{n+1}^\text{ext} + \mathcal{A}_{n+1}^\text{int}.
\end{equation}
In addition to the set of on-shell momenta
$\{p\}=\{p_1,\,...\,,p_i,\,...\,,p_n\}$ we
define the sets of momenta $\{p\}_i=\{p_1,\,...\,,p_i-k,\,...\,,p_n\}$
that are adapted to emission from line $i$.  Taking all particles
apart from the emitted photon to be incoming (but ignoring the complex
conjugation of the polarisation vector $\epsilon$) allows us to write
the soft expansion of $\mathcal{A}_{n+1}^\text{ext}$ as
\begin{subequations}
\begin{align}
    \mathcal{A}_{n+1}^\text{ext}
    \,=\,& \sum_i Q_i 
    \frac{\Gamma^\text{ext}(\{p\}_i)(\slashed{p}_i-\slashed{k}+m)\gamma^\mu u(p_i) \epsilon_\mu(k)}{2 k\cdot p_i} \\
    \,=\,& \sum_i Q_i \Big(
    \frac{\epsilon\cdot p_i}{k\cdot p_i}\Gamma^\text{ext}(\{p\}_i)
    - \frac{\Gamma^\text{ext}(\{p\}_i)\slashed{k}\slashed{\epsilon}}{2 k\cdot p_i}
    \Big) u(p_i) \\
    \wideeq{k\sim\xi}& \sum_i Q_i \Big(
    \frac{1}{\xi}\frac{\epsilon\cdot p_i}{k\cdot p_i} \Gamma^\text{ext}(\{p\})
    -\frac{\epsilon\cdot p_i}{k\cdot p_i} k\cdot\frac{\partial}{\partial p_i} \Gamma^\text{ext}(\{p\}) 
    -\frac{\Gamma^\text{ext}(\{p\})\slashed{k}\slashed{\epsilon}}{2k\cdot p_i}
    \Big) u(p_i)
    + \mathcal{O}(\xi),
\end{align}
\end{subequations}
Since $\{p\}$ satisfies the radiative momentum conservation $\sum_i
p_i = k$ this is not a strict expansion in $\xi$.
Following~\cite{Adler:1966gc} we can make the above split gauge
invariant (up to subleading power) via the modification
\begin{subequations}
\begin{align} \label{eq:ext-ginv}
    &\mathcal{A}^\text{ext}_{n+1} 
    \to \mathcal{A}^\text{I}_{n+1}
    \equiv \epsilon\cdot A^\text{I}_{n+1}
    = \mathcal{A}^\text{ext}_{n+1} 
    + \sum_i Q_i \epsilon\cdot\pder{p_i} \Gamma^\text{ext}(\{p\}) u(p_i), \\
    &\mathcal{A}^\text{int}_{n+1} 
    \to \mathcal{A}^\text{II}_{n+1}
    \equiv \epsilon\cdot A^\text{II}_{n+1}
    = \mathcal{A}^\text{int}_{n+1} 
    - \sum_i Q_i \epsilon\cdot\pder{p_i} \Gamma^\text{ext}(\{p\}) u(p_i).
\end{align}
\end{subequations}
Indeed, $k\cdot  A^\text{I}_{n+1} \sim \mathcal{O}(\xi^2)$. The
leading contributions in $\xi$ vanish due to $\sum_i Q_i=0$ and the
subleading contributions cancel between the two terms of the last
expression in \eqref{eq:ext-ginv}. Because the full amplitude is gauge
invariant we also have  $k\cdot  A^\text{II}_{n+1} \sim
\mathcal{O}(\xi^2)$. Following the reasoning in~\cite{Adler:1966gc},
we observe that the leading $\mathcal{O}(\xi^0)$ term in
$A^\text{II}_{n+1}$ must be independent of $k$ due to the lack of
$1/k$ poles in $\mathcal{A}_{n+1}^\text{int}$. Combined with the Ward
identity this directly implies that $\mathcal{A}^\text{II}_{n+1} \sim
\mathcal{O}(\xi)$. As a consequence, the soft expansion of the total amplitude can be
written as
\begin{subequations}
\begin{equation}
    \mathcal{A}_{n+1}^{(0)} = \sum_i Q_i \Big(
    \frac{1}{\xi}\frac{\epsilon\cdot p_i}{k\cdot p_i}\Gamma^\text{ext}(\{p\})
    -\frac{\Gamma^\text{ext}(\{p\})\slashed{k}\slashed{\epsilon}}{2k\cdot p_i}
    -\big[\epsilon\cdot D_i \Gamma^\text{ext}(\{p\})\big]
    \Big) u(p_i)+\mathcal{O}(\xi)
\end{equation}
with the LBK operator
\begin{equation}
    D_i^\mu = \frac{p_i^\mu}{k\cdot p_i}k
    \cdot\frac{\partial}{\partial p_i}
    - \frac{\partial}{\partial p_{i,\mu}}.
\end{equation}
\end{subequations}
Squaring the amplitude, summing over spins and polarisations, and
using the identity
\begin{align}
    \frac{(\slashed{p_i}+m)\slashed{\epsilon}\slashed{k}
    +\slashed{k}\slashed{\epsilon}(\slashed{p_i}+m)}{2k\cdot p_i}
    =\frac{\epsilon\cdot p_i}{k\cdot p_i}\slashed{k}-\slashed{\epsilon}
    =\epsilon\cdot D_i (\slashed{p_i}+m)
\end{align}
then yields for the matrix element
\begin{equation}\label{eq:lbk}
    \mathcal{M}_{n+1}^{(0)}(\{p\},k)
    =\sum_{ij} Q_i Q_j \Big(
    -\frac{1}{\xi^2} \frac{p_i\cdot p_j}{(k\cdot p_i)(k\cdot p_j)}
    + \frac{1}{\xi} \frac{p_j\cdot D_i}{k\cdot p_j}
    \Big) \mathcal{M}_n^{(0)}(\{p\})
    + \mathcal{O}(\xi^0).
\end{equation}
This shows that not only the leading term in the soft expansion is
related to the non-radiative matrix element but that this is also true
at subleading power at tree level. However, the non-radiative matrix
element in \eqref{eq:lbk} is evaluated with a set of momenta $\{p\}$
that does not satisfy momentum conservation. This is unproblematic for
the tree-level matrix elements considered here. If, on the other hand,
loop corrections are taken into account
(Section~\ref{sec:lbk_oneloop}) this significantly complicates the
evaluation of the corresponding integrals. In this case a different
formulation of the LBK theorem is helpful. To this end, we reabsorb
the first term of the LBK operator into the matrix element to undo
the expansion and write
\begin{equation}\label{eq:lbk_interm}
    \mathcal{M}_{n+1}^{(0)}(\{p\},k)
    = -\sum_{ij} Q_i Q_j \Big(
    \frac{1}{\xi^2} \frac{p_i\cdot p_j}{(k\cdot p_i)(k\cdot p_j)}
    +\frac{1}{\xi}\frac{1}{k\cdot p_j}p_j\cdot \pder{p_i} \Big)
    \mathcal{M}_n^{(0)}(\{p\}_i) + \mathcal{O}(\xi^0).
\end{equation}
Since $\{p\}_i$ satisfies momentum conservation we can now express the
non-radiative matrix element in terms of invariants
\begin{align}
    \mathcal{M}_{n}^{(0)}(\{p\}_i)
    = \mathcal{M}_{n}^{(0)} (\{s\}_i,\{m^2\}_i)
\end{align}
with $\{s\}_i=\big\{s(\{p\}_i,\{m^2\}_i)\big\}$ and
$\{m^2\}_i=\{p_1^2=m_1^2,\,...\,,(p_i-k)^2,\,...\,,p_n^2=m_n^2\,\}$.
The corresponding expansion in $k$ can then be written  as
\begin{align} \nonumber
    \lefteqn{\mathcal{M}_{n}^{(0)} (\{s\}_i,\{m^2\}_i)
    =} \\ & \Bigg(
    1-\xi
    \sum_L \Big( k\cdot\pder[s_L]{p_i}+2k\cdot p_i\pder[s_L]{m_i^2}\Big)\pder{s_L}
    -\xi\, 2k\cdot p_i\pder{m_i^2}
    \Bigg)
    \mathcal{M}_{n}^{(0)} (\{s\},\{m^2\})
    + \mathcal{O}(\xi^2),
    \label{eq:deriv1}
\end{align}
where the sum $L$ is over the set of independent invariants
$\{s\}=\big\{s(\{p\},\{m^2\})\big\}$ expressed in terms of the momenta
$\{p\}$ and the on-shell masses $\{m^2\}$.  Similarly, we can write
\begin{align} \nonumber
    \lefteqn{p_j\cdot \pder{p_i} \mathcal{M}_{n}^{(0)} (\{s\}_i,\{m^2\}_i)
    = } \\ & \Bigg(
    \sum_L\Big(p_j\cdot\pder[s_L]{p_i}+2p_i\cdot p_j\pder[s_L]{m_i^2} \Big) \pder{s_L}
    + 2p_i\cdot p_j\pder{m_i^2}
    \Bigg)
    \mathcal{M}_{n}^{(0)} (\{s\},\{m^2\})
    + \mathcal{O}(\xi).
    \label{eq:deriv2}
\end{align}
Inserting~\eqref{eq:deriv1} and~\eqref{eq:deriv2}
into~\eqref{eq:lbk_interm}, all derivatives with respect to the masses
cancel and we obtain the simple formulation of the LBK theorem in
terms of invariants
\begin{subequations}\label{eq:lbk-inv}
\begin{align}
    \mathcal{M}_{n+1}^{(0)}(\{p\},k)
    = \Big(
    \frac{1}{\xi^2}\eik 
    + \frac{1}{\xi} \sum_{ij} Q_i Q_j \frac{p_j\cdot\tilde{D}_i}{k\cdot p_j}
    \Big) \mathcal{M}_{n}^{(0)} (\{s\},\{m^2\})
    + \mathcal{O}(\xi^0)
\end{align}
with the modified LBK operator
\begin{align}\label{eq:lbkop}
    \tilde{D}_i^\mu 
    = \sum_L \Big(
    \frac{p_i^\mu}{k\cdot p_i} k\cdot \pder[s_L]{p_i}-\pder[s_L]{p_{i,\mu}}
    \Big) \pder{s_L}.
\end{align}
\end{subequations}
The advantage of \eqref{eq:lbk-inv} over \eqref{eq:lbk} is that
conventional one-loop techniques can be applied in this case. We
emphasise that the choice of $\{s\}=\big\{s(\{p\},\{m^2\})\big\}$ is
ambiguous since the momenta $\{p\}$ do not satisfy momentum
conservation. This is however not an issue as long as the same
definition is used in the calculation of the derivatives $\partial
s_L/\partial p_i^\mu$. The above formula can therefore be conveniently used to analytically compute the soft limit of tree-level matrix elements up to subleading power. An alternative approach that is particularly suitable for the numerical evaluation of the LBK theorem was recently presented in~\cite{Bonocore:2021cbv}.

The above formula assumes all particles apart from the photon to be
incoming. In the case of outgoing particles the corresponding momentum
$p$ has to be replaced with $-p$. In particular, this also implies
$\partial/\partial p^\mu\to-\partial/\partial p^\mu$.

\subsection{One-loop generalisation of the LBK theorem}
\label{sec:lbk_oneloop}

The derivation of the previous section cannot be applied one-to-one in
the presence of loop corrections due to contributions from regions
where the loop momentum is small ($\ell \sim \xi$). In this soft
region the power counting used to derive the LBK theorem is no longer
valid. In particular, individual contributions from internal emissions
already contribute at leading power. Fortunately, the method of
regions~\cite{Beneke:1997zp} can be used to disentangle the 
soft contribution from the hard
one. For the latter the LBK formula~\eqref{eq:lbk} still holds. The
soft contribution will be evaluated in a generic way in the following
sections. The collinear regions, on the other hand, all vanish due to the absence of collinear scales in the soft limit as defined in Section~\ref{sec:notation}. The combination of hard and soft contributions, that will be given in Section~\ref{sec:lbk_full}, therefore generalises the LBK theorem to one loop.

\subsubsection{General considerations regarding the soft contribution}

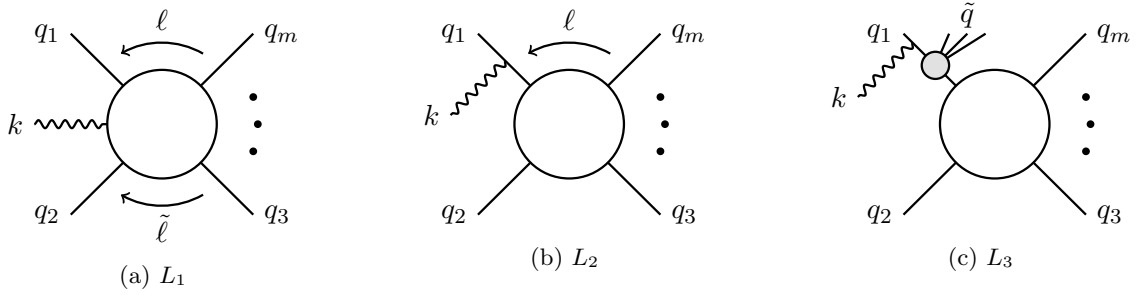
\begin{figure}
    \centering
    \subfloat[$L_1$]{
        \begin{tikzpicture}[scale=1.2,baseline={(0,0)}]
            \draw[line width=.3mm] (-1,-1) node[left]{$q_2$} -- (0,0) -- (1,-1) node[right]{$q_3$};
    \draw[line width=.3mm] (-1,+1) node[left]{$q_1$} -- (0,0) -- (1,+1) node[right]{$q_m$};
    \draw[line width=.3mm]  [tightphoton]  (-1.4,0) node[left]{$k$} -- (-0.6,0);
    \draw[line width=.3mm,fill=white] (0,0) circle (0.6);
    
    \draw[line width=.3mm,fill=black] (1,0.3) circle (0.03);
    \draw[line width=.3mm,fill=black] (1.05,0) circle (0.03);
    \draw[line width=.3mm,fill=black] (1,-0.3) circle (0.03);
    
    \centerarc [line width=0.3mm, ->](0,0)(60:120:0.9);
    \node[] at (0,1.15) {$\ell$};
    \centerarc [line width=0.3mm, ->](0,0)(300:240:0.9);
    \node[] at (0,-1.15) {$\tilde{\ell}$};
    
    \end{tikzpicture}
    \label{fig:loop1}
    }
    \hspace{1cm}
    \subfloat[$L_2$]{
        \begin{tikzpicture}[scale=1.2,baseline={(0,0)}]
            \draw[line width=.3mm] (-1,-1) node[left]{$q_2$} -- (0,0) -- (1,-1) node[right]{$q_3$};
    \draw[line width=.3mm] (-1,+1) node[left]{$q_1$} -- (0,0) -- (1,+1) node[right]{$q_m$};
    \draw[line width=.3mm]  [tightphoton]  (-.7,.7) -- (-1.3,.1) node[left]{$k$};
    \draw[line width=.3mm,fill=white] (0,0) circle (0.6);
    
    \draw[line width=.3mm,fill=black] (1,0.3) circle (0.03);
    \draw[line width=.3mm,fill=black] (1.05,0) circle (0.03);
    \draw[line width=.3mm,fill=black] (1,-0.3) circle (0.03);
    
    \centerarc [line width=0.3mm, ->](0,0)(60:120:0.9);
    \node[] at (0,1.15) {$\ell$};
    
    \end{tikzpicture}
    \label{fig:loop2}
    }
    \hspace{1cm}
    \subfloat[$L_3$]{
        \begin{tikzpicture}[scale=1.2,baseline={(0,0)}]
          
    \draw[line width=.3mm] (-1,-1) node[left]{$q_2$} -- (0,0) -- (1,-1) node[right]{$q_3$};
    \draw[line width=.3mm] (-1,+1) node[left]{$q_1$} -- (0,0) -- (1,+1) node[right]{$q_m$};
    \draw[line width=.3mm,fill=white] (0,0) circle (0.6);
    \draw[line width=.3mm]  [tightphoton]  (-.9,.9) -- (-1.5,.3) node[left]{$k$};
    
    \draw[line width=.3mm] (-.65,.65) -- (-.5,1);
    \draw[line width=.3mm] (-.65,.65) -- (-.3,1);
    \draw[line width=.3mm] (-.65,.65) -- (-.1,1);
    \draw[line width=.3mm,fill=stuff] (-.65,.65) circle (0.15);
    
    \node[] at (-.3,1.2) {$\tilde{q}$};
    
    \draw[line width=.3mm,fill=black] (1,0.3) circle (0.03);
    \draw[line width=.3mm,fill=black] (1.05,0) circle (0.03);
    \draw[line width=.3mm,fill=black] (1,-0.3) circle (0.03);
    \end{tikzpicture}
    \label{fig:loop3}
    }
\caption{Classification of one-loop integrals encountered in the
calculation of radiative QED amplitudes. The legs that connect to the
loop can either be on shell or off shell.}
\label{fig:loops}
\end{figure}

In order to systematically analyse possible origins of the soft
contribution we classify the one-loop integrals as illustrated in
Figure~\ref{fig:loops} where the momenta $q_i$ can be off shell
(internal) or on shell (external). The circle symbolises the one-loop
integral associated with the one-particle irreducible (1PI) part of a
particular Feynman diagram. The first class, $L_1$, includes
$(m+1)$-point integrals from diagrams where the photon is directly
attached to this 1PI part. Class $L_2$ includes $m$-point integrals
from diagrams where the photon is attached to a leg that directly
connects to the 1PI part with momentum $q_1-k$. As we will see, the
treatment of these integrals depends on whether the momentum of the
adjacent leg $q_m$ is on shell or off shell. Finally, for integrals of
the type $L_3$ the photon is attached indirectly to the $m$-point 1PI
part such that the momentum flowing into the loop integral is
$q_1-k+\tilde{q}$ with a non-zero $\tilde{q}$. 

For integrals to have a non-vanishing soft contribution the momentum
routing has to be chosen such that the loop momentum $\ell$ is aligned
with a photon propagator. All other choices lead only to linear
propagators in the soft momentum expansion and therefore vanish as a
consequence of the residue theorem. There can thus be at most as many
soft regions as the number of photons in the loop. However, most of
them yield scaleless integrals and vanish in dimensional
regularisation. This is in particular the case for all possible
routings of $L_3$. The presence of the momentum $\tilde{q}$ allows to
set $\ell=0$ for the soft contribution in all propagators except for
the photon propagator with momentum $\ell$. Hence, loop integrals of
the form $L_3$ do not contribute to the soft region. For the second
class, on the other hand, there is one non-vanishing soft contribution
indicated by the momentum routing in Figure~\ref{fig:loop2} if the
corresponding internal propagator is given by a photon and if in
addition $q_1$ is on shell. In the case where $q_m$ is off shell the
soft expansion starts at $\mathcal{O}(\xi)$ and is given by
\begin{equation}
   I_{L_2}(q_m^2\neq m_m^2) = \int \frac{\text{d}^d \ell}{(2\pi)^d} \frac{1}{
      [\ell^2]\ [2\ell\cdot q_1-2k\cdot q_1]\ [(q_1+q_2)^2-m_2^2]\ ...\ [q_m^2-m_m^2]}
      .
\end{equation}
For on-shell $q_m$ the leading integral reads instead
\begin{equation}
   I_{L_2}(q_m^2 = m_m^2) = \int \frac{\text{d}^d \ell}{(2\pi)^d} \frac{1}{
      [\ell^2]\ [2\ell\cdot q_1-2k\cdot q_1]\ [(q_1+q_2)^2-m_2^2]\ ...\ [-2\ell\cdot q_m]}
\end{equation}
which already contributes at $\mathcal{O}(\xi^0)$. Finally, the first
class of loop integrals $L_1$ gives rise to up to two non-vanishing
soft contributions given by the two momentum routings $\ell$ and
$\tilde{\ell}$ in Figure~\ref{fig:loop1} if the corresponding
propagators are photons. The integral for routing $\ell$
\begin{align}
   \lefteqn{I_{L_1} = } \notag\\&
   \int \frac{\text{d}^d \ell}{(2\pi)^d} \frac{1}{
      [\ell^2]\ [2\ell\cdot q_1+q_1^2-m_1^2] \ [2\ell\cdot q_1-2k\cdot q_1+q_1^2-m_1^2]\ ...\ [-2\ell\cdot q_m+q_m^2-m_m^2]}
\end{align}  
is only non-zero if $q_1$ is on shell and it starts to contribute at
$\mathcal{O}(\xi^{-1})$ if $q_m$ is on shell and at
$\mathcal{O}(\xi^0)$ otherwise. The analogous statements hold for the
$\tilde{\ell}$ momentum routing.

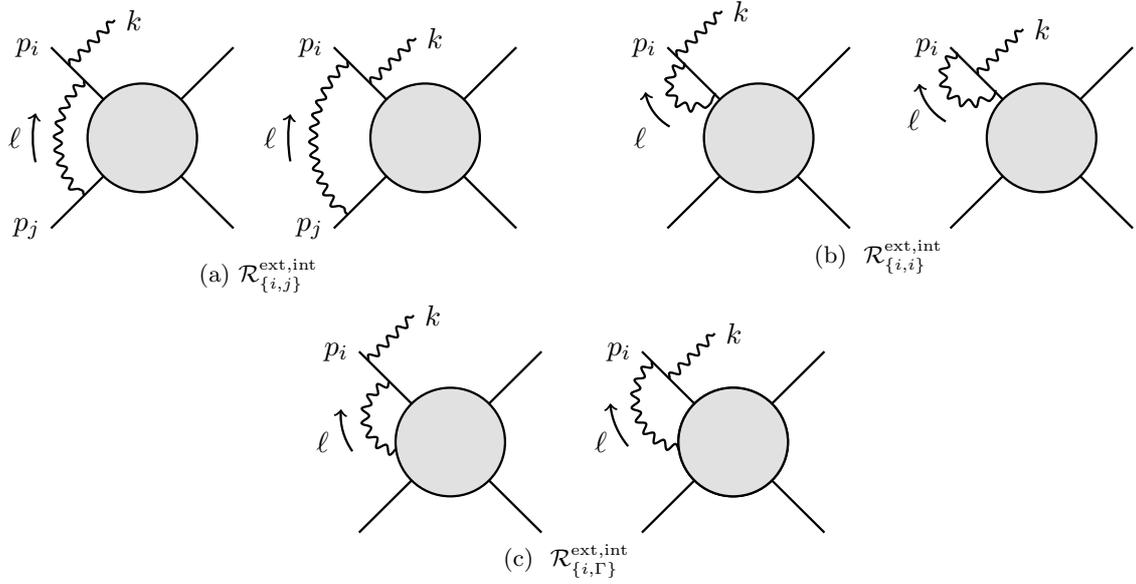
\begin{figure}
    \centering
    \subfloat[$\mathcal{R}_{\{i,j\}}^{\text{ext},\text{int}}$]{
        \begin{tikzpicture}[scale=1.2,baseline={(0,0)}]
         
    \draw[line width=.3mm] (-1,-1) node[left]{$p_j$} -- (0,0) -- (1,-1);
    \draw[line width=.3mm] (-1,+1) node[left]{$p_i$}  -- (0,0) -- (1,+1);
    \draw[line width=.3mm]  [tightphoton]  (-.8,.8) -- (-.3,1.3) node[right]{$k$};
    \draw[line width=.3mm,fill=stuff] (0,0) circle (0.6);
    
    \centerarc [line width=0.3mm, tightphoton](0,0)(135:225:.9);
    \centerarc [line width=0.3mm, ->](0,0)(193:167:1.2);
    \node[] at (-1.4,0) {$\ell$};
    
    \draw[line width=.3mm] (-1+3.1,-1) node[left]{$p_j$} -- (0+3.1,0) -- (1+3.1,-1);
    \draw[line width=.3mm] (-1+3.1,+1) node[left]{$p_i$}  -- (0+3.1,0) -- (1+3.1,+1);
    \draw[line width=.3mm]  [tightphoton]  (-.6+3.1,.6) -- (-.1+3.1,1.1) node[right]{$k$};
    \draw[line width=.3mm,fill=stuff] (0+3.1,0) circle (0.6);
    
    \centerarc [line width=0.3mm, tightphoton](0+3.1,0)(135:225:1.2);
    \centerarc [line width=0.3mm, ->](0+3.1,0)(190:170:1.5);
    \node[] at (-1.7+3.1,0) {$\ell$};
    \end{tikzpicture}
    \label{fig:softdiag1}
    }
    \hspace{1cm}
    \subfloat[
    $\mathcal{R}_{\{i,i\}}^{\text{ext},\text{int}}$]{
        \begin{tikzpicture}[scale=1.2,baseline={(0,0)}]
            \draw[line width=.3mm] (-1,-1) -- (0,0) -- (1,-1);
    \draw[line width=.3mm] (-1,+1) node[left]{$p_i$}  -- (0,0) -- (1,+1);
    \draw[line width=.3mm]  [tightphoton]  (-.9,.9) -- (-.4,1.4) node[right]{$k$};
    \draw[line width=.3mm,fill=stuff] (0,0) circle (0.6);
    
    \centerarc [line width=0.3mm, tightphoton](-.7,.6)(115:335:0.25);
    \centerarc [line width=0.3mm, ->](-.7,.6)(240:190:0.55);
    \node[] at (-1.3,0) {$\ell$};
    
    \draw[line width=.3mm] (-1+3.1,-1) -- (0+3.1,0) -- (1+3.1,-1);
    \draw[line width=.3mm] (-1+3.1,+1) node[left]{$p_i$}  -- (0+3.1,0) -- (1+3.1,+1);
    \draw[line width=.3mm]  [tightphoton]  (-.7+3.1,.7) -- (-.2+3.1,1.2) node[right]{$k$};
    \draw[line width=.3mm,fill=stuff] (0+3.1,0) circle (0.6);

    \centerarc [line width=0.3mm, tightphoton](-.75+3.1,.7)(125:325:.3);
    \centerarc [line width=0.3mm, ->](-.75+3.1,.7)(240:190:.6);
    \node[] at (-1.4+3.1,.2) {$\ell$};
    \end{tikzpicture}
    \label{fig:softdiag2}
    }
    \hspace{1cm}
    \subfloat[
    $\mathcal{R}_{\{i,\Gamma\}}^{\text{ext},\text{int}}$]{
        \begin{tikzpicture}[scale=1.2,baseline={(0,0)}]
         
    \draw[line width=.3mm] (-1,-1) -- (0,0) -- (1,-1);
    \draw[line width=.3mm] (-1,+1) node[left]{$p_i$}  -- (0,0) -- (1,+1);
    \draw[line width=.3mm]  [tightphoton]  (-.9,.9) -- (-.4,1.4) node[right]{$k$};
    
    \centerarc [line width=0.3mm, tightphoton](-.5,.3)(115:360:.4);
    \centerarc [line width=0.3mm, ->](-.5,.3)(215:175:.7);
    \draw[line width=.3mm,fill=stuff] (0,0) circle (0.6);
    \node[] at (-1.4,0) {$\ell$};
    
    \draw[line width=.3mm] (-1+3.1,-1) -- (0+3.1,0) -- (1+3.1,-1);
    \draw[line width=.3mm] (-1+3.1,+1) node[left]{$p_i$}  -- (0+3.1,0) -- (1+3.1,+1);
    \draw[line width=.3mm]  [tightphoton]  (-.7+3.1,.7) -- (-.2+3.1,1.2) node[right]{$k$};
    \draw[line width=.3mm,fill=stuff] (0+3.1,0) circle (0.6);
    
    \centerarc [line width=0.3mm, tightphoton](-.5+3.1,.5)(135:360:.55);
    \centerarc [line width=0.3mm, ->](-.5+3.1,.5)(220:185:.85);
    \node[] at (-1.45+3.1,0) {$\ell$};
    \draw[line width=.3mm,fill=stuff] (0+3.1,0) circle (0.6);
    \end{tikzpicture}
    \label{fig:softdiag3}
    }
\caption{Diagrammatic classification of the soft contributions each
corresponding to a particular momentum routing of a Feynman diagram.}
\label{fig:softdiags}
\end{figure}

The above reasoning allows to represent every possible soft
contribution according to the three pairs of diagrams shown in
Figure~\ref{fig:softdiags} where the external legs are now all on
shell. Every $\mathcal{R}_{\{e,a\}}^\text{int, ext}$ corresponds to an
amplitude with a specific choice of the momentum routing where the
soft contribution does not vanish. The labels for emission and
absorption $\{e,a\}$ take on the values $1\ldots n$ or $\Gamma$. The
superscript $\text{int, ext}$ indicates whether the photon $k$ is
attached internally or externally. In the former (latter) case we are
dealing with integrals of the type $L_1$ ($L_2$). As mentioned in
connection with $L_1$, it is possible that one amplitude contributes to
two soft representations. Taking $e=i$ and assuming $p_i$ to be
incoming, we can write the corresponding expressions generically as 
\begin{subequations}\label{eq:softrep}
\begin{align}
    &i \mathcal{R}_{\{e,a\}}^\text{ext}
    = \frac{
    Q_i^2 \Gamma_{\{e,a\}}^\mu(\slashed{\ell}+\slashed{p_i}-\slashed{k}+m)
    \gamma_\mu(\slashed{p_i}-\slashed{k}+m)\slashed{\epsilon}u(p_i)
    }{
    -2 k\cdot p_i [\ell^2][\ell^2+2\ell\cdot(p_i-k)-2k\cdot p_i]
    }, \\
    &i \mathcal{R}_{\{e,a\}}^\text{int}
    = \frac{
    Q_i^2 \Gamma_{\{e,a\}}^\mu(\slashed{\ell}+\slashed{p_i}-\slashed{k}+m)
    \slashed{\epsilon}(\slashed{\ell}+\slashed{p_i}+m)\gamma_\mu u(p_i)
    }{
    [\ell^2][\ell^2+2\ell\cdot(p_i-k)-2k\cdot p_i][\ell^2+2\ell\cdot p_i]
    }.
\end{align}
\end{subequations}
All terms related to the emission from leg  $e=i$ and the soft photon
propagator are given explicitly in \eqref{eq:softrep}. The vertex and
fermion propagator related to the absorption is common to
$\mathcal{R}_{\{e,a\}}^\text{ext}$ and
$\mathcal{R}_{\{e,a\}}^\text{int}$ and is included in
$\Gamma_{\{e,a\}}^\mu$. This implies the scalings $\Gamma_{\{i,j\}}^\mu
\sim \Gamma_{\{i,i\}}^\mu \sim \xi^{-1}$ and
$\Gamma_{\{i,\Gamma\}}^\mu \sim \xi^0$. We then write the expansion in the soft region
of the sum of the diagram pairs as
\begin{subequations}
\begin{align}
    i \mathcal{R}_{\{e,a\}}^\text{soft}
    \wideeq{k\sim\xi} &\frac{1}{\xi}S_{\{e,a\}}^\text{LP}+S_{\{e,a\}}^\text{NLP}+\mathcal{O}(\xi) \\
    \,=\,& \frac{1}{\xi} \Big(
    S_{\{e,a\}}^{\text{LP},\text{ext}} + S_{\{e,a\}}^{\text{LP},\text{int}} \Big) + S_{\{e,a\}}^{\text{NLP},\text{ext}} + S_{\{e,a\}}^{\text{NLP},\text{int}}+\mathcal{O}(\xi),
\end{align}
\end{subequations}
with the leading and subleading power terms denoted by
$S_{\{e,a\}}^\text{LP}$ and $S_{\{e,a\}}^\text{NLP}$, respectively.
Based on the previously discussed power counting of the integrals
$L_1$ and $L_2$ we can deduce that
$S_{\{i,\Gamma\}}^\text{LP}=S_{\{i,\Gamma\}}^{\text{LP},\text{ext}}=S_{\{i,\Gamma\}}^{\text{LP},\text{int}}=0$.

\subsubsection{Vanishing of the soft contribution at leading power}\label{sec:softLP}

The leading-power soft contribution to~\eqref{eq:softrep} is given by
\begin{subequations}
\begin{align}
    &S_{\{e,a\}}^{\text{LP},\text{ext}}
    = \frac{
    Q_i^2 \Gamma_{\{e,a\}}^\mu(\slashed{p_i}+m)
    \gamma_\mu(\slashed{p_i}+m)\slashed{\epsilon}u(p_i)
    }{
    -4 k\cdot p_i [\ell^2][\ell\cdot p_i-k\cdot p_i]
    }
    =
    -\frac{
    Q_i^2 \Gamma_{\{e,a\}}^\mu u(p_i) p_{i,\mu} p_i\cdot\epsilon
    }{
    [\ell^2][\ell\cdot p_i-k\cdot p_i][k\cdot p_i]
    }
    , \\
    &S_{\{e,a\}}^{\text{LP},\text{int}}
    = \frac{
    Q_i^2 \Gamma_{\{e,a\}}^\mu (\slashed{p_i}+m)
    \slashed{\epsilon}(\slashed{p_i}+m)\gamma_\mu u(p_i)
    }{
    4[\ell^2][\ell\cdot p_i-k\cdot p_i][\ell\cdot p_i]
    }
    =
    +\frac{
    Q_i^2 \Gamma_{\{e,a\}}^\mu u(p_i) p_{i,\mu} p_i\cdot\epsilon
    }{
    [\ell^2][\ell\cdot p_i-k\cdot p_i][\ell\cdot p_i]
    }.
\end{align}
\end{subequations}
The propagators can be brought to the same form with the partial
fraction identity
\begin{equation}
    \frac{1}{[\ell\cdot p_i-k\cdot p_i][\ell\cdot p_i]}
    =\frac{1}{k\cdot p_i} \Big(
    \frac{1}{[\ell\cdot p_i-k\cdot p_i]}-\frac{1}{[\ell\cdot p_i]} \Big)
\end{equation}
where the second term in the curly brackets can be neglected up to
scaleless integrals. We then see immediately that
\begin{equation}
    S_{\{e,a\}}^\text{LP}=S_{\{e,a\}}^{\text{LP},\text{ext}}+S_{\{e,a\}}^{\text{LP},\text{int}}=0.
\end{equation}
We therefore arrive at the known result that the eikonal approximation in QED
does not receive genuine loop corrections. Furthermore, this also
shows that $S_{\{i,\Gamma\}}^\text{NLP}=0$ since it effectively
corresponds to a leading-power contribution. The third class of soft
contributions, $\mathcal{R}_{\{i,\Gamma\}}^\text{soft}$, can therefore
be omitted in the following discussion.

\subsubsection{Soft contribution at subleading power}

At subleading power there are contributions in~\eqref{eq:softrep} from
either the higher-order expansion of propagators (denominator) or from
numerator terms proportional to $\ell$ or $k$. We therefore write
\begin{equation}
    S_{\{e,a\}}^\text{NLP} 
    = S_{\{e,a\}}^{(\text{NLP},\text{D})}+S_{\{e,a\}}^{(\text{NLP},\text{N})}
    = \Big(
       S_{\{e,a\}}^{(\text{NLP},\text{D}),\text{ext}}+S_{\{e,a\}}^{(\text{NLP},\text{D}),\text{int}}
    \Big) + \Big(
       S_{\{e,a\}}^{(\text{NLP},\text{N}),\text{ext}}+S_{\{e,a\}}^{(\text{NLP},\text{N}),\text{int}}
    \Big).
\end{equation}

For the denominator type the leading-power cancellation of
Section~\ref{sec:softLP} occurs if propagators other than
$[\ell^2+2\ell\cdot(p_i-k)-2 k\cdot p_i]$ or $[\ell^2+2\ell\cdot p_i]$
are expanded. Furthermore, expansion in $\ell^2$ of these two propagators results only in linear propagators.
Consequently, we only have to consider the expansion in $\ell\cdot k$ of the
propagator $[\ell^2+2\ell\cdot(p_i-k)-2 k\cdot p_i]$ in~\eqref{eq:softrep}. Using
partial fraction then yields the simple contribution
\begin{equation}
    S_{\{e,a\}}^{(\text{NLP},\text{D})} 
    = S_{\{e,a\}}^{(\text{NLP},\text{D}),\text{ext}}+S_{\{e,a\}}^{(\text{NLP},\text{D}),\text{int}} 
    = -\frac{
    Q_i^2 \Gamma_{\{e,a\}}^\mu u(p_i) p_{i,\mu} p_i\cdot\epsilon\, \ell\cdot k
    }{
    (k\cdot p_i)^2 [\ell^2][\ell\cdot p_i-k\cdot p_i]
    }\ .
\end{equation}
The numerator type can be written as
\begin{subequations}
\begin{equation}
    S_{\{e,a\}}^{(\text{NLP},\text{N})} 
    = \frac{Q_i^2 \Gamma_{\{e,a\}}^\mu}{4 k\cdot p_i[\ell^2][\ell\cdot p_i-k\cdot p_i]}
    \big(T^\text{ext}_\mu+K^\text{ext}_\mu+T^\text{int}_\mu+K^\text{int}_\mu \big)u(p_i)
\end{equation}
with
\begin{align}
    T^\text{ext}_\mu
    &= -(\slashed{\ell}-\slashed{k})\gamma_\mu(\slashed{p}_i+m)\slashed{\epsilon}, \\
    K^\text{ext}_\mu
    &= (\slashed{p}_i+m)\gamma_\mu\slashed{k}\slashed{\epsilon}, \\
    T^\text{int}_\mu
    &= (\slashed{\ell}-\slashed{k})\slashed{\epsilon}(\slashed{p}_i+m)\gamma_\mu, \\
    K^\text{int}_\mu
    &= (\slashed{p}_i+m)\slashed{\epsilon}\slashed{\ell}\gamma_\mu.
\end{align}
\end{subequations}
Due to various cancellations among $T_\mu^{\text{ext},\text{int}}$ and
$K_\mu^{\text{ext},\text{int}}$ we obtain the simple result
\begin{equation}
    S_{\{e,a\}}^{(\text{NLP},\text{N})}
    = \frac{Q_i^2 \Gamma_{\{e,a\}}^\mu u(p_i)}{k\cdot p_i [\ell^2][\ell\cdot p_i-k\cdot p_i]}
    \big( k_\mu p_i\cdot\epsilon-\epsilon_\mu k\cdot p_i+p_{i,\mu} \ell\cdot\epsilon \big),
\end{equation}
where we have used the replacement $\ell\cdot p_i\to k\cdot p_i$ in
the numerator which holds up to scaleless integrals.

To make further progress at this point we need to treat
$S_{\{i,j\}}^\text{NLP}$ and $S_{\{i,i\}}^\text{NLP}$ separately in
order to specify the form of $\Gamma_{\{e,a\}}^\mu$. In the case where
the photon is reabsorbed by the emitting leg, i.e.
$S_{\{i,i\}}^\text{NLP}$, we have
\begin{equation}
    \Gamma_{\{i,i\}}^\mu u(p_i)
    = -\frac{Q_i \Gamma^{(0)} (\slashed{p_i}-\slashed{k}+m)\gamma^\mu u(p_i)}{-2 k\cdot p_i}
    \wideeq{k\sim\xi}  \frac{1}{\xi} Q_i \mathcal{A}_n^{(0)} \frac{p_i^\mu}{k\cdot p_i} + \mathcal{O}(\xi^0)
\end{equation}
where $\mathcal{A}_n^{(0)}=\Gamma^{(0)} u(p_i)$ corresponds to the
non-radiative tree-level amplitude. In this case, we further have the
simple Passarino-Veltman decomposition
\begin{equation}
    \ell^\rho \to \frac{\ell\cdot p_i}{m_i^2} p_i^\rho
\end{equation}
where we can again replace $\ell\cdot p_i$ with $k\cdot p_i$. It is
then straightforward to see that
\begin{equation}
    S_{\{i,i\}}^\text{NLP}=S_{\{i,i\}}^{(\text{NLP},\text{D})}+S_{\{i,i\}}^{(\text{NLP},\text{N})} = 0 .
\end{equation}
Hence, diagrams where the loop corrects only the emitting leg do not
contribute at subleading power.

In the case of $S_{\{i,j\}}^\text{NLP}$, on the other hand, we find
for an incoming particle $p_j$ that
\begin{equation}
    \Gamma_{\{i,j\}}^\mu u(p_i) 
    \wideeq{k\sim\xi} - \frac{1}{\xi} Q_j \mathcal{A}_n^{(0)} \frac{p_j^\mu}{[-\ell\cdot p_j]} + \mathcal{O}(\xi^0).
\end{equation}
This in turn results after the tensor decomposition
\begin{align}
    \ell^\rho \to
    \frac{\ell \cdot p_j\, p_i\cdot p_j-\ell \cdot p_i\, m_j^2}{(p_i \cdot p_j)^2-m_i^2 m_j^2}\,p_i^\rho
    + \frac{\ell \cdot p_i\, p_i\cdot p_j-\ell \cdot p_j\, m_i^2}{(p_i \cdot p_j)^2-m_i^2 m_j^2}\,p_j^\rho
\end{align}
in the only non-vanishing subleading power contribution of the form
\begin{subequations}\label{eq:amp_soft}
\begin{equation}
    S_{\{i,j\}}^\text{NLP}
    \equiv S_{\{i,j\}}^\text{NLP}(p_i,p_j,Q_i,Q_j)
    = - Q_i^2 Q_j (i\mathcal{A}_n^{(0)})
     \Big( \frac{p_i\cdot\epsilon}{k\cdot p_i}-\frac{p_j\cdot\epsilon}{k\cdot p_j} \Big)
     S(p_i,p_j,k),
\end{equation}
where we have defined the function
\begin{equation}\label{eq:softfunc}
    S(p_i,p_j,k)
    = \frac{m_i^2 k\cdot p_j}{\big((p_i\cdot p_j)^2-m_i^2 m_j^2\big)k\cdot p_i}
      \Big( p_i\cdot p_j I_1(p_i,k)+m_j^2 k\cdot p_i I_2(p_i,p_j,k) \Big).
\end{equation}
The analytic results for the integrals
\begin{align}
    I_1(p_i,k) &= i \int \frac{\text{d}^d \ell}{(2\pi)^d} \frac{1}{[\ell^2+i\delta][\ell\cdot p_i-k\cdot p_i+i\delta]}, \label{eq:softinta}\\
    I_2(p_i,p_j,k) &= i \int \frac{\text{d}^d \ell}{(2\pi)^d} \frac{1}{[\ell^2+i\delta][-\ell\cdot p_j+i\delta][\ell\cdot p_i-k\cdot p_i+i\delta]}\label{eq:softintb}
\end{align}
\end{subequations}
can be found in Appendix~\ref{sec:softints}. The causal $+i\delta$
prescription is given explicitly in the above integrals.

The result~\eqref{eq:amp_soft} is also valid for incoming
antiparticles with the overall sign difference parametrised by the
fermion charges $Q_i$ and $Q_j$. The total soft contribution can thus
be obtained by summing the above expression over all external charged
fermions, i.e.
\begin{equation}\label{eq:amp_soft_full}
    \mathcal{A}_{n+1}^{(1),\text{soft}}
    = \frac{1}{\xi} \sum_{i\neq j} S_{\{i,j\}}^\text{NLP}(p_i,p_j,Q_i,Q_j) + \mathcal{O}(\xi^0).
\end{equation}
The corresponding expression for the matrix element can be obtained by
interfering with the eikonal approximation of the tree-level
amplitude. The resulting formula is given in the following section.

\subsubsection{One-loop extension of the LBK theorem}\label{sec:lbk_full}

Based on the previous discussion we find that the one-loop correction
to a generic radiative process in QED in the
limit where the emitted photon becomes soft satisfies the expansion
\begin{subequations}\label{eq:lbk_oneloop}
\begin{equation}
    \mathcal{M}_{n+1}^{(1)}(\{p\},k)
    \wideeq{k\sim\xi} 
    \mathcal{M}_{n+1}^{(1),\text{hard}} + \mathcal{M}_{n+1}^{(1),\text{soft}}
\end{equation}
with
\makeatletter
\newcommand{\pushleft}[1]{\ifmeasuring@#1\else\omit$\displaystyle#1$\hfill\fi\ignorespaces}
\makeatother
\begin{alignat}{3}
    \label{eq:lbk_oneloop_hard}
    \mathcal{M}_{n+1}^{(1),\text{hard}}
    &=& \phantom{\frac1\xi}\sum_l\sum_i Q_i Q_l \Big(
    - \frac{1}{\xi^2}
    \frac{p_i\cdot p_l}{(k\cdot p_i)(k\cdot p_l)}
    + \frac{1}{\xi}
    \frac{p_l\cdot \tilde{D}_i}{k\cdot p_l}
    \Big)
    \mathcal{M}_n^{(1)} (\{s\},\{m^2\})
    &+& \mathcal{O}(\xi^0), \\
    \label{eq:lbk_oneloop_soft}
    \mathcal{M}_{n+1}^{(1),\text{soft}}
    &=& \pushleft{\frac{1}{\xi}\sum_l \sum_{i \neq j} Q_i^2 Q_j Q_l
    \Big( \frac{p_i\cdot p_l}{(k\cdot p_i)(k\cdot p_l)} 
    - \frac{p_j\cdot p_l}{(k\cdot p_j)(k\cdot p_l)} \Big) 2
    S(p_i,p_j,k)}\notag \\
    && \times  \mathcal{M}_{n}^{(0)}(\{s\},\{m^2\}) &+& \mathcal{O}(\xi^0).
\end{alignat}
\end{subequations}
This is the generalisation of the LBK theorem at one loop. We
emphasize that the above result assumes all particles to be incoming.
For outgoing particles one can simply replace the corresponding
momentum $p$ with $-p$. Furthermore, the LBK operator $\tilde{D}_i$
and the function $S(p_i,p_j,k)$ are defined in~\eqref{eq:lbkop}
and~\eqref{eq:softfunc}, respectively. 

A conceptual illustration of the factorisation formula~\eqref{eq:lbk_oneloop} is shown in Figure~\ref{fig:soft_fac}. Contributions with hard and soft origin are depicted in green and orange, respectively. The first two diagrams on the r.h.s correspond to the hard sector given by~\eqref{eq:lbk_oneloop_hard}. The factorisation of~\eqref{eq:lbk_oneloop_soft} into a universal soft function - connecting three external legs simultaneously - and the non-radiative matrix element is illustrated in the third diagram.
Based on this, a naive extrapolation to higher orders in perturbation theory is possible by interpreting Figure~\ref{fig:soft_fac} as an all-order statement. First of all, this would imply that the LBK operator yields the hard contribution also beyond one loop. More interestingly, however, it would significantly constrain the mixed hard-soft structure. At two loops, for example, the hard-soft region would be fixed through objects that already enter in~\eqref{eq:lbk_oneloop}. In particular, it would correspond to~\eqref{eq:lbk_oneloop_soft} with $\mathcal{M}_n^{(0)}\to\mathcal{M}_n^{(1)}$. The only new contribution in the factorisation formula would therefore be the two-loop soft function corresponding to the purely soft region.

\begin{figure}
    \centering
    \begin{tikzpicture}[scale=1,baseline={(1,0)}]
    	    \draw[line width=.3mm] (-1,-1)-- (0,0) -- (1,-1);
    \draw[line width=.3mm]  (-1,+1) -- (0,0) -- (1,+1);
    \draw[line width=.3mm]  [tightphoton] (.2,+1.2) node[right]{$k$} -- (0,0);
    
    \draw[line width=.3mm]  [fill=stuff] (0,0) circle (0.6);
    
    \node at (2,0) {$\wideeq{k\sim \xi}$};
    
    \draw[line width=.3mm]  (-1+4.5,-1) -- (0+4.5,0) -- (1+4.5,-1);
    \draw[line width=.3mm]  (-1+4.5,+1) -- (0+4.5,0) -- (1+4.5,+1);
    %\draw[line width=.3mm]  [tightphoton, hard] (-1.2+4.5,0) -- (-0.8+4.5,0.8);
    %\draw[line width=.3mm]  [tightphoton, hard] (-1.2+4.5,0) -- (-.8+4.5,-.8);
    \centerarc [line width=0.3mm,tightphoton, hard](0+4.5,0)(135:225:1.1);
    \draw[line width=.3mm]  [fill=hard] (-1.2+4.5,0) circle (0.35) node[]{$\eik$};
    \draw[line width=.3mm]  [fill=hard] (0+4.5,0) circle (0.45) node[]{$\mathcal{M}_n$};
    
    \node at (6,0) {$+$};
    
    \draw[line width=.3mm]  (-1+8.5,-1) -- (0+8.5,0) -- (1+8.5,-1);
    \draw[line width=.3mm]  (-1+8.5,+1) -- (0+8.5,0) -- (1+8.5,+1);
    %\draw[line width=.3mm]  [tightphoton, hard] (-1.2+8.5,0) -- (-0.8+8.5,0.8);
    %\draw[line width=.3mm]  [tightphoton, hard] (-1.2+8.5,0) -- (-.8+8.5,-.8);
    \centerarc [line width=0.3mm,tightphoton, hard](0+8.5,0)(135:225:1.1);
    \draw[line width=.3mm]  [fill=hard] (-1.2+8.5,0) circle (0.35) node[]{$D$};
    \draw[line width=.3mm]  [fill=hard] (0+8.5,0) circle (0.45) node[]{$\mathcal{M}_n$};
    
    \node at (10,0) {$+$};
    
    \draw[line width=.3mm]  (-1+12.5,-1) -- (0+12.5,0) -- (1+12.5,-1);
    \draw[line width=.3mm]  (-1+12.5,+1) -- (0+12.5,0) -- (1+12.5,+1);
    %\draw[line width=.3mm]  [tightphoton, soft] (-1.2+12.5,0) -- (-0.9+12.5,0.9);
    %\draw[line width=.3mm]  [tightphoton, soft] (-1.2+12.5,0) -- (-.8+12.5,-.8);
    \centerarc [line width=0.3mm,tightphoton, soft](0+12.5,0)(135:225:1.1);
    \centerarc [line width=0.3mm,tightphoton, soft](12.5,-.5)(160:62:1.1);
    %\draw[line width=.3mm]  [tightphoton, soft] (-1.2+12.5,0) -- (.9+12.5,.9);
    \draw[line width=.3mm]  [fill=hard] (0+12.5,0) circle (0.45) node[]{$\mathcal{M}_n$};
    \draw[line width=.3mm]  [fill=soft] (-1.2+12.5,0) circle (0.35) node[]{$S$};

    \node at (14,0) {$+$};
    \node at (15,0) {$\mathcal{O}(\xi^0)$};
    \end{tikzpicture}
    \caption{Schematic illustration of the soft factorisation
    of~\eqref{eq:lbk_oneloop} at subleading power and at one loop.}
    \label{fig:soft_fac}
\end{figure}

\section{Collinear factorisation at leading power}
\label{sec:collinear} 

In this section we consider the leading-power behaviour of radiative
amplitudes in the limit where the emitted photon ($k$) becomes
collinear to a light on-shell fermion ($p$). This configuration is
governed by the scale hierarchy
\begin{equation}
    k\cdot p \sim m^2 \sim \lambda^2 \ll \{s\} \sim \lambda^0,
\end{equation}
where $\{s\}$ denotes all other scales in the process. As mentioned in the
introduction it was shown a long time ago that the leading-power
contribution to the tree-level amplitude factorises in this limit into
a process-independent splitting function multiplying the non-radiative
amplitude with radiative
kinematics~\cite{Baier:1973ms,Berends:1981uq,Kleiss:1986ct,Dittmaier:1999mb}.
It is the goal of this section to present the one-loop generalisation
of this formula. We first start in Section~\ref{subsec:coll_tree} by
reproducing the tree-level derivation from~\cite{Baier:1973ms} and
then discuss the one-loop extension in Section~\ref{subsec:coll_loop}.
The final factorisation formula is then given
in~\eqref{eq:collfac_isr} for initial-state radiation (ISR) and
in~\eqref{eq:collfac_fsr} for final-state radiation (FSR).

\subsection{Collinear factorisation at tree level}
\label{subsec:coll_tree}

Contrary to the soft limit, care has to be taken in the collinear case
when treating the gauge dependence of the emitted photon. Only axial
gauge, where the sum over photon polarisations is given by
\begin{equation}\label{eq:axialgauge}
    \sum_{\text{pol}} 
    \epsilon_\mu(k) \epsilon_\nu(k)
    = -g_{\mu\nu}+\frac{k_\mu r_\nu+k_\nu r_\mu}{k\cdot r},
\end{equation}
does not mix up the power counting of individual diagrams. At tree
level a convenient choice for the gauge vector $r$ is
$\bar{k}\equiv(E_k,-\vec{k})$. At leading power in the collinear limit
we therefore only need to consider diagrams where the photon is
emitted from the collinear fermion leg. Restricting the discussion for
the moment to ISR, we have
\begin{equation}
    \mathcal{A}^{(0)}_{n+1} 
    \wideeq{k\cdot p\sim \lambda^2} \begin{tikzpicture}[scale=.8,baseline={(0,-.1)}]
        
   \draw[line width=.3mm]  (-1,0) node[left]{\footnotesize${p}$} -- (1,0);
    \draw[line width=0.3mm,photon]  (-.3,0) -- (0,.65) node[right]{\footnotesize${k}$};
    
    \draw[line width=.3mm]  [fill=stuff] (1,0) circle (0.45) node[black]{$\Gamma_0$};
       \end{tikzpicture} + \mathcal{O}(\lambda^{0})
    = - Q \Gamma_0(p-k)
    \frac{\slashed{p}-\slashed{k}+m}{-2 k\cdot p}
    \gamma^\mu u(p) \epsilon_\mu(k) + \mathcal{O}(\lambda^{0}).
\end{equation}
We then write the fermion
propagator in terms of quasi-real spinors~\cite{Baier:1973ms} with
energy
\begin{equation}
    E_{p-k} 
    \equiv \sqrt{(\vec{p}-\vec{k})^2+m^2}
    =E_{p}-E_k + \lambda^2\frac{k\cdot p}{E_{p}-E_k}+\mathcal{O}(\lambda^4)
\end{equation}
as
\begin{subequations}
\begin{align}
    \frac{\slashed{p}-\slashed{k}+m}{-2 k\cdot p}
    & = \frac{1}{2 E_{p-k}} \sum_s \Big(
        \frac{u^s(p_{ik}) \bar{u}^s(p_{ik})}{E_{p}-E_k-E_{p-k}}
      + \frac{v^s(\bar{p}_{ik}) \bar{v}^s(\bar{p}_{ik})}{E_{p}-E_k+E_{p-k}}
    \Big) \\
    & \wideeq{k\cdot p\sim\lambda^2}
    -\frac{1}{\lambda^2}\frac{1}{2 k\cdot p}\sum_s 
        u^s(p_{ik}) \bar{u}^s(p_{ik})
      + \mathcal{O}(\lambda^{-1})
\end{align}
\end{subequations}
with $p_{ik} = (E_{p-k},\vec{p}-\vec{k})$ and $\bar{p}_{ik}
=(E_{p-k},-\vec{p}+\vec{k})$. It is then straightforward to derive the
factorised result for the matrix element
\begin{equation}\label{eq:collfac_treelevel}
    \mathcal{M}^{(0)}_{n+1} 
    \wideeq{k\cdot p \sim \lambda^2} 
    \frac{1}{\lambda^2} J^{(0)}_\text{ISR}(x,m)\ \mathcal{M}^{(0)}_n(p-k,m=0) + \mathcal{O}(\lambda^{-1}),
\end{equation}
where the tree-level splitting function for ISR in its standard
form~\cite{Dittmaier:1999mb} reads
\begin{equation}
    J^{(0)}_\text{ISR}(x,m)
    = \frac{Q^2}{x (k\cdot p)} \Big(
    \frac{1+x^2}{1-x}-\frac{x\,m^2}{k\cdot p} \Big), \quad
    x = \frac{E_p-E_k}{E_p}.
\end{equation}
The analogous derivation for FSR yields
\begin{equation}
    \mathcal{M}^{(0)}_{n+1} 
    \wideeq{k\cdot p \sim \lambda^2} 
    \frac{1}{\lambda^2} J^{(0)}_\text{FSR}(z,m)\ \mathcal{M}^{(0)}_n(p+k,m=0) + \mathcal{O}(\lambda^{-1})
\end{equation}
with
\begin{equation}
    J^{(0)}_\text{FSR}(z,m)
    = \frac{Q^2}{k\cdot p} \Big(
    \frac{1+z^2}{1-z}-\frac{m^2}{k\cdot p} \Big), \quad
    z = \frac{E_p}{E_p+E_k}.
\end{equation}
Alternatively, $J_\text{FSR}^{(0)}(z,m)$ can be derived from
$J_\text{ISR}^{(0)}(x,m)$ via the crossing relation $p\to -p$, i.e. by
replacing $k\cdot p \to -k\cdot p$ and $x\to z^{-1}$.

\subsection{Collinear factorisation at one loop}
\label{subsec:coll_loop}

\begin{figure}
    \centering
    \subfloat[$\mathcal{M}_1$]{
        \begin{tikzpicture}[scale=1,baseline={(0,0)}]
            
    \draw[line width=.3mm]  (-1.75,2.5) -- (1.75,2.5);
    \draw[line width=.3mm]  (-1.75,0) -- (1.75,0);
    \draw[line width=.3mm]  (-1.75,-.1) -- (1.75,-.1);
    
    \centerarc [line width=0.3mm,photon](-0.875,2.5)(0:180:0.6)
    
    \draw [photon, line width=0.3mm] (-1.7,2.5) to[out=-70,in=-10] (1.5,1.3);
    \draw[line width=.3mm]  [white,fill=white] (1.7,1.3) circle (0.2);
    \draw[line width=.3mm]  [white,fill=white] (0,1.4) circle (0.2);
    \draw [photon, line width=0.3mm] (0,2.5)--(0,0);
    
    \draw[line width=.3mm,dashed]  (2,3) -- (2,-.5);
    
    \draw[line width=.3mm]  (2.25,2.5) -- (5.75,2.5);
    \draw[line width=.3mm]  (2.25,0) -- (5.75,0);
    \draw[line width=.3mm]  (2.25,-.1) -- (5.75,-.1);
    
    \draw [photon, line width=0.3mm] (5.25,2.5) to[out=-100,in=0] (2.4,1.3);
    \draw[line width=.3mm]  [white,fill=white] (4,1.3) circle (0.2);
    \draw[line width=.3mm]  [white,fill=white] (2.3,1.3) circle (0.2);
    \draw [photon, line width=0.3mm] (4,2.5)--(4,0);
    \end{tikzpicture}
    \label{mue_diag1}
    }
    \subfloat[$\mathcal{M}_2$]{
        \begin{tikzpicture}[scale=1,baseline={(0,0)}]
            

    \draw[line width=.3mm]  (-1.75,2.5) -- (1.75,2.5);
    \draw[line width=.3mm]  (-1.75,0) -- (1.75,0);
    \draw[line width=.3mm]  (-1.75,-.1) -- (1.75,-.1);
    
    \centerarc [line width=0.3mm,photon](-0.875,2.5)(0:180:0.6)
    
    \draw [photon, line width=0.3mm] (-1,2.5) to[out=-70,in=-10] (1.5,1.3);
    \draw[line width=.3mm]  [white,fill=white] (1.7,1.3) circle (0.25);
    \draw[line width=.3mm]  [white,fill=white] (0,1.7) circle (0.2);
    \draw [photon, line width=0.3mm] (0,2.5)--(0,0);
    
    \draw[line width=.3mm,dashed]  (2,3) -- (2,-.5);
    
    \draw[line width=.3mm]  (2.25,2.5) -- (5.75,2.5);
    \draw[line width=.3mm]  (2.25,0) -- (5.75,0);
    \draw[line width=.3mm]  (2.25,-.1) -- (5.75,-.1);
    
    \draw [photon, line width=0.3mm] (5.25,2.5) to[out=-100,in=0] (2.4,1.3);
    \draw[line width=.3mm]  [white,fill=white] (4,1.3) circle (0.2);
    \draw[line width=.3mm]  [white,fill=white] (2.3,1.3) circle (0.2);
    \draw [photon, line width=0.3mm] (4,2.5)--(4,0);
    \end{tikzpicture}
    \label{mue_diag2}
    }\\
    \subfloat[$\mathcal{M}_3$]{
        \begin{tikzpicture}[scale=1,baseline={(0,0)}]
            \draw[line width=.3mm]  (-1.75,2.5) -- (1.75,2.5);
    \draw[line width=.3mm]  (-1.75,0) -- (1.75,0);
    \draw[line width=.3mm]  (-1.75,-.1) -- (1.75,-.1);
    
    \centerarc [line width=0.3mm,photon](0,2.5)(0:180:0.6)
    
    \draw [photon, line width=0.3mm] (-1.3,2.5) to[out=-70,in=-10] (1.5,1.3);
    \draw[line width=.3mm]  [white,fill=white] (1.7,1.3) circle (0.2);
    \draw[line width=.3mm]  [white,fill=white] (0,1.7) circle (0.2);
    \draw [photon, line width=0.3mm] (0,2.5)--(0,0);
    
    \draw[line width=.3mm,dashed]  (2,3) -- (2,-.5);
    
    \draw[line width=.3mm]  (2.25,2.5) -- (5.75,2.5);
    \draw[line width=.3mm]  (2.25,0) -- (5.75,0);
    \draw[line width=.3mm]  (2.25,-.1) -- (5.75,-.1);
    
    \draw [photon, line width=0.3mm] (5.25,2.5) to[out=-100,in=0] (2.4,1.3);
    \draw[line width=.3mm]  [white,fill=white] (4,1.3) circle (0.2);
    \draw[line width=.3mm]  [white,fill=white] (2.3,1.3) circle (0.2);
    \draw [photon, line width=0.3mm] (4,2.5)--(4,0);
    \end{tikzpicture}
    \label{mue_diag3}
    }
    \subfloat[$\mathcal{M}_4$]{
        \begin{tikzpicture}[scale=1,baseline={(0,0)}]
            \draw[line width=.3mm]  (-1.75,2.5) -- (1.75,2.5);
    \draw[line width=.3mm]  (-1.75,0) -- (1.75,0);
    \draw[line width=.3mm]  (-1.75,-.1) -- (1.75,-.1);
    
    \centerarc [line width=0.3mm,photon](0,2.5)(0:180:0.6)
    
    \draw [photon, line width=0.3mm] (-0.35,2.5) to[out=-70,in=-10] (1.5,1.3);
    \draw[line width=.3mm]  [white,fill=white] (1.7,1.3) circle (0.2);
    \draw[line width=.3mm]  [white,fill=white] (0,2) circle (0.2);
    \draw [photon, line width=0.3mm] (0,2.5)--(0,0);
    
    \draw[line width=.3mm,dashed]  (2,3) -- (2,-.5);
    
    \draw[line width=.3mm]  (2.25,2.5) -- (5.75,2.5);
    \draw[line width=.3mm]  (2.25,0) -- (5.75,0);
    \draw[line width=.3mm]  (2.25,-.1) -- (5.75,-.1);
    
    \draw [photon, line width=0.3mm] (5.25,2.5) to[out=-100,in=0] (2.4,1.3);
    \draw[line width=.3mm]  [white,fill=white] (4,1.3) circle (0.2);
    \draw[line width=.3mm]  [white,fill=white] (2.3,1.3) circle (0.2);
    \draw [photon, line width=0.3mm] (4,2.5)--(4,0);
    \end{tikzpicture}
    \label{mue_diag4}
    }
\caption{Interference terms that contribute at leading power in the limit where the emitted photon becomes
collinear to the initial-state electron.}
\label{fig:mue_diags}
\end{figure}
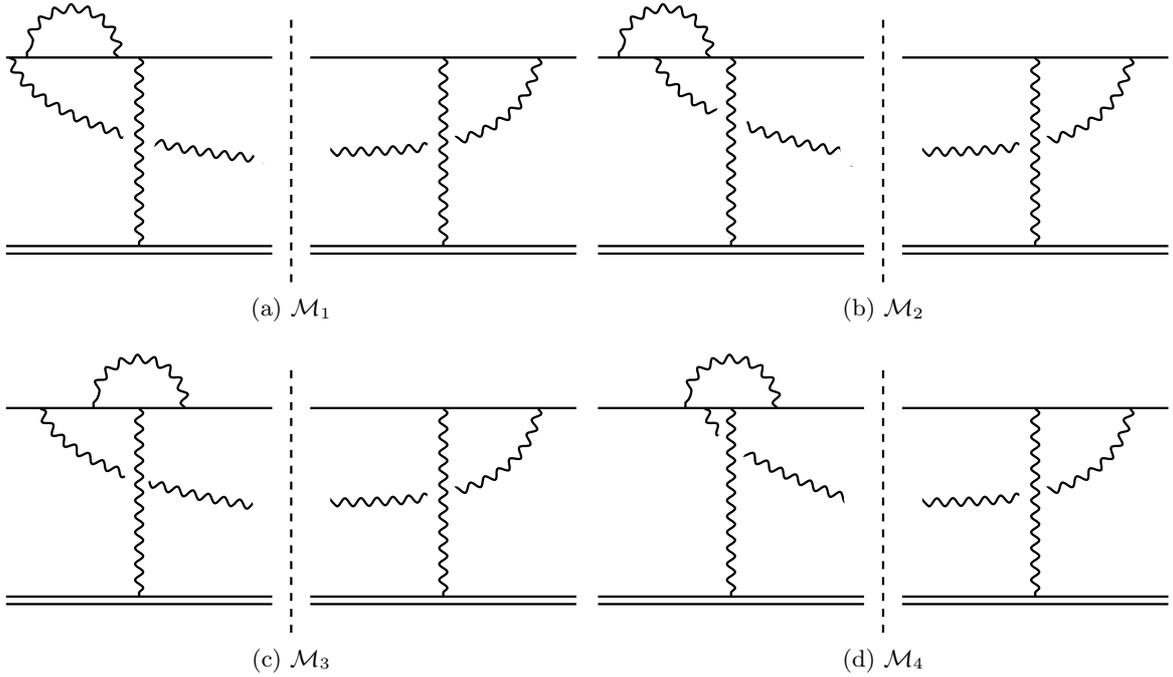

The quasi-real electron method from the previous section does not work
anymore if loop corrections are taken into account due to
non-factorisable diagrams where the photon is emitted from a loop.
However, the method of regions can be applied in this case to
disentangle universal contributions from collinear degrees of freedom
from the process-dependent hard part. It is therefore possible to
determine the splitting function based on a specific process. To this
end we consider muon-electron scattering, i.e. 
\begin{equation}
    e(p_1)\mu(q_1)\to e(p_2)\mu(q_2)\gamma(k)
\end{equation}
and calculate the small-mass collinear limit of the one-loop
corrections to the electron line. The scale hierarchy for ISR then
reads
\begin{equation}
    k\cdot p_1,\ p_1^2=p_2^2=m^2 \sim \lambda^2 \ll q_1^2=q_2^2=M^2,\ \{s\} \sim \lambda^0.
\end{equation}
Working at leading power and in axial gauge, we only need to take the
four interference terms shown in Figure~\ref{fig:mue_diags} into
account. In particular, we have used the convenient choice $r=p_2$ for
the gauge vector. This choice is allowed since $p_2^2=m^2$ is
small. Note that $r=p_1$ would not be permissible in this case since
the small numerator $k\cdot p_1$ in~\eqref{eq:axialgauge} would mix up
the power counting.

The momenta $p_i=E_i(1,\vec{n}_i\beta_i)$ of the energetic electrons
can be decomposed into large and small components via the set of
light-cone basis vectors $\{n_i=(1,\vec{n}_i)/\sqrt{2},
\bar{n}_i=(1,-\vec{n}_i)/\sqrt{2}\}$ which allows us to write any
momentum as
\begin{subequations}
\begin{align}
    p_j
    &= (n_i\cdot p_j) \bar{n}_i+(\bar{n}_i\cdot p_j)n_i+ p_{j,(\perp,i)} \\
    &= p_j^{(+,i)} + p_j^{(-,i)} + p_j^{(\perp,i)} \\
    &= (n_i\cdot p_j, \bar{n}_i\cdot p_j, p_{j,\perp})_i.
\end{align}
\end{subequations}
In the case of the energetic particle $p_i$ this takes the form of the
desired decomposition
\begin{equation}
    p_i = \big(E_i(1-\beta_i)/\sqrt{2},E_i(1+\beta_i)/\sqrt{2},p_{i,\perp}\big)_i \sim (\lambda^2,1,\lambda)_i
\end{equation}
where the scaling can be deduced from
\begin{equation}
    p_j^2 = 2 p_j^{(+,i)}\cdot p_j^{(-,i)}+\big(p_j^{(\perp,i)}\big)^2=m_j^2\sim\lambda^2.
\end{equation}
Applying the light-cone decomposition to the external momenta in our
process we find
\begin{subequations}
\begin{align}
    &p_1 \sim k \sim (\lambda^2,1,\lambda)_1\sim (1,1,1)_2, \\
    &p_2 \sim (1,1,1)_1\sim (\lambda^2,1,\lambda)_2, \\
    &q_1 \sim q_2 \sim (1,1,1)_1 \sim (1,1,1)_2.
\end{align}
\end{subequations}

Based on the achieved disentanglement of scales it is possible to
identify the contributing momentum regions. In order to do so it is
helpful to use the formulation of the method of regions in the alpha
parameter representation~\cite{Smirnov:1999bza} automatised in the
public code {\tt asy.m}~\cite{Jantzen:2012mw}. The following four
regions are then found to contribute to the individual interference
terms $\mathcal{M}_i$:
\begin{subequations}
\begin{alignat}{3}
    &\mbox{hard:}                  \quad &\ell& \sim (1,1,1)_1               \ \ &\sim&\ \  (1,1,1)_2 \\
    &\mbox{$p_1$-collinear:}       \quad &\ell& \sim (\lambda^2,1,\lambda)_1 \ \ &\sim&\ \  (1,1,1)_2 \\
    &\mbox{$p_2$-collinear:}       \quad &\ell& \sim (1,1,1)_1               \ \ &\sim&\ \  (\lambda^2,1,\lambda)_2\\
    &\mbox{$p_2$-ultra-collinear:} \quad &\ell& \sim (1,1,1)_1               \ \ &\sim&\ \  (\lambda^4,\lambda^2,\lambda^3)_2
\end{alignat}
\end{subequations}
The terms that correct the incoming electron line, i.e.
$\mathcal{M}_1$ and $\mathcal{M}_2$, get only contributions from the
$p_1$-collinear region. Furthermore, at leading power the hard region
only contributes to the factorisable diagram $\mathcal{M}_3$.  Since
we can apply the quasi-real electron method in this case it follows
immediately that
\begin{equation}\label{eq:fac_hard}
    \mathcal{M}_{n+1}^{(1),\text{hard}} 
    = \frac{1}{\lambda^2}J^{(0)}_\text{ISR}(x,m) \mathcal{M}^{(1)}_n(p_1-k,m=0) + \mathcal{O}(\lambda^{-1}).
\end{equation}
In addition to the hard region, all other three scalings contribute to
$\mathcal{M}_3$. In the case of $\mathcal{M}_4$, on the other hand,
only the $p_1$-collinear and $p_2$-ultra-collinear regions are present
at leading power. As can be expected, the unphysical ultra-collinear
region cancels between $\mathcal{M}_3$ and $\mathcal{M}_4$. We are
then left with the two collinear contributions that turn out to
factorise according to
\begin{subequations}
\begin{align}
    \mathcal{M}^{(1),p_1\text{-coll}}_{n+1} 
    &= \frac{1}{\lambda^2} J^{(1)}_\text{ISR}(x,m)\ \mathcal{M}_n^{(0)}(p_1-k,m=0) + \mathcal{O}(\lambda^{-1}), \\
    \mathcal{M}^{(1),p_2\text{-coll}}_{n+1} 
    &= \frac{1}{\lambda^2} Z^{(1)}(m) \ J^{(0)}_\text{ISR}(x,m)\ \mathcal{M}_n^{(0)}(p_1-k,m=0) + \mathcal{O}(\lambda^{-1}).
\end{align}
\end{subequations}
Apart from the interference terms $\mathcal{M}_i$ we also need to take
into account mass and wave function renormalisation. All counterterms
connected to the heavy particles (muon) enter in~\eqref{eq:fac_hard}
in the renormalisation of the non-radiative massless one-loop matrix
element $\mathcal{M}_n^{(1)}$. The counterterms for the emitting
electron, on the other hand, renormalise the one-loop splitting
function $J^{(1)}_\text{ISR}$, while the ones for the other light
particle (outgoing electron) contributes to $Z^{(1)}$. The
renormalised results for $J^{(1)}_\text{ISR}$ and $Z^{(1)}$ are
given in Appendix~\ref{sec:splitfunc}. 

The factor $Z^{(1)}$ corresponds to the one-loop massification constant
of~\cite{Engel:2018fsb}. Massification is a method to efficiently
determine the leading mass effects of an amplitude solely based on the massless result.
All mass terms that are not polynomially suppressed are recovered in this way.
It is the universality of the aforementioned massification constant that makes such a reconstruction possible. The $p_2$-collinear contribution therefore takes the leading-order mass effects of the outgoing electron into account. The one-loop splitting function
$J^{(1)}_\text{ISR}$ contains both the corresponding
mass terms as well as leading-power corrections due to the collinear
emission.

A non-trivial check for the above result is the behaviour in the soft limit
\begin{equation}
    J_\text{ISR}^{(1)} \to Z^{(1)} \eik_\text{coll}, \quad
    J_\text{ISR}^{(0)} \to \eik_\text{coll},
\end{equation}
where $\eik_\text{coll}$ corresponds to the eikonal factor in the
collinear limit. We therefore get the expected form of the matrix element in the
collinear-soft limit given by
\begin{align}
    \mathcal{M}_{n+1}^{(1)} 
    &\to \eik_\text{coll} 
    \Big(\mathcal{M}^{(1)}(p_1,m=0)
        +2 Z^{(1)}(m)\mathcal{M}_n^{(0)}(p_1,m=0)
    \Big) = \eik_\text{coll} \mathcal{M}^{(1),\text{massified}}_{n}.
\end{align}

\begin{figure}
    \centering
    \begin{tikzpicture}[scale=1,baseline={(1,0)}]
    	    \draw[line width=.3mm] (-1,-1) -- (0,0) -- (1,-1);
    \draw[line width=.3mm]  (-1,+1) node[left]{$p_j$} -- (0,0) -- (1,+1)node[right]{$p_i$};
    \draw[line width=.3mm]  [tightphoton] (.2,+1.2) node[right]{$k$} -- (0,0);
    
    \draw[line width=.3mm]  [fill=stuff] (0,0) circle (0.6);

    \draw[line width=.3mm]  (-1+4,-1) -- (0+4,0) -- (1+4,-1);
    \draw[line width=.3mm]  (-1+4,+1) -- (0+4,0) -- (1+4,+1);
    
    \node at (2,0) {$\wideeq{k \cdot p_i, m_j \sim \lambda^2}$};
    \node at (6,0) {$\quad +\quad \mathcal{O}(\lambda^{-1})$};
    
    \draw[line width=.3mm]  [fill=hard] (0+4,0) circle (0.45);
    
    \draw[line width=.3mm]  [fill=col1,rotate around={atan(-1):(0+4,0)} ] (0+4,0.9) ellipse (0.2 and 0.35) node[]{$J$};
    \draw[line width=.3mm]  [fill=col4,rotate around={atan(1):(0+4,0)} ] (0+4,0.9) ellipse (0.2 and 0.35)node[]{$Z$};
    \end{tikzpicture}
    \caption{Schematic illustration of the collinear factorisation
    formulas of~\eqref{eq:collfac_isr} and~\eqref{eq:collfac_fsr}.}
    \label{fig:coll_fac}
\end{figure}
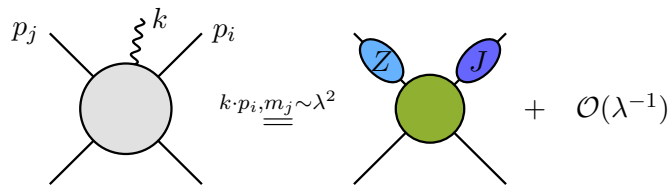

As already mentioned previously the collinear contributions are
expected to be process independent. Thus, one-loop diagrams for
muon-electron scattering other than those shown in
Figure~\ref{fig:mue_diags} should not lead to such contributions.  
We have explicitly checked that this is the case due to a cancellation
between diagram pairs that are related (up to a sign) through the
crossing $q_1\leftrightarrow-q_2$. The only additional contribution is
therefore the hard one originating from factorisable diagrams that
trivially exhibit the factorising structure of~\eqref{eq:fac_hard}.
If, on the other hand, we take the muon to be light as well, i.e.
$M^2\sim m^2\sim \lambda^2$, there are two additional collinear
contributions with exactly the same structure as for the outgoing
electron
\begin{equation}
    \mathcal{M}_{n+1}^{(1),q_1\text{-coll}}=
    \mathcal{M}_{n+1}^{(1),q_2\text{-coll}}
    =\frac{1}{\lambda^2} Z^{(1)}(M) J_\text{ISR}^{(0)}(x,m) \mathcal{M}_n^{(0)}(p_1-k,m=0,M=0)
    +\mathcal{O}(\lambda^{-1})
\end{equation}
consistent with the expectation based on massification. We are thus
lead to the main result of this section that at one loop can be
written through the factorisation formula
\begin{subequations}\label{eq:collfac_isr}
\begin{equation}
    \mathcal{M}_{n+1} \wideeq{k\cdot p_i, m_j^2 \sim \lambda^2}
    \frac{1}{\lambda^2}J_\text{ISR}(x,m_i) \Bigg( \prod_{j\neq i} Z(m_j) \Bigg) \mathcal{M}_n(p_i-k,m_i=0,m_j=0)
    + \mathcal{O}(\lambda^{-1}),
\end{equation}
where we have defined the all order quantities
\begin{align}
    &J_\text{ISR} = J_\text{ISR}^{(0)} + J_\text{ISR}^{(1)} + \mathcal{O}(\alpha^3), \\
    &Z = 1 + Z^{(1)} + \mathcal{O}(\alpha^2).
\end{align}
\end{subequations}
In \eqref{eq:collfac_isr} the product is over all external fermion
lines $j\neq i$ with a small mass $m_j^2\sim\lambda^2$.  Furthermore,
the same calculation with only minor modifications can also be applied
to the case of FSR yielding the analogous formula
\begin{equation}\label{eq:collfac_fsr}
    \mathcal{M}_{n+1} \wideeq{k\cdot p_i, m_j^2 \sim \lambda^2}
    \frac{1}{\lambda^2}J_\text{FSR}(z,m_i) \Bigg(\prod_{j\neq i} Z(m_j)\Bigg) \mathcal{M}_n(p_i+k,m_i=0,m_j=0)
    + \mathcal{O}(\lambda^{-1}).
\end{equation}
A schematic illustration of these factorisation formulas is given in
Figure~\ref{fig:coll_fac}. Furthermore, the corresponding expressions
for $J_\text{ISR}$, $J_\text{FSR}$, and $Z$ can be found in
Appendix~\ref{sec:splitfunc}. As expected, we find that the ISR and
FSR splitting functions are related via crossing symmetry.

It is useful to compare our findings to the corresponding
factorisation formula for massless fermions that can be extracted from
the QCD results of~\cite{Bern:1994zx}. Suppressing the separation into
ISR and FSR, the massless collinear limit can be written as
\begin{align}
    \mathcal{M}_{n+1} \wideeq{k\cdot p_i \to 0} \bar{J}(y) \mathcal{M}_n,
\end{align}
where the only process-independent contribution $\bar{J}$ comes from
the collinear fermion ($p_i$) and $y\in\{x,z\}$. The corresponding expressions at tree level and at one loop are given in~Appendix~\ref{sec:splitfunc}. For massive particles, on the other
hand, every light fermion contributes an additional factor in the
factorisation formula taking into account the corresponding small-mass
effects. This results in the more complex collinear structure
of~\eqref{eq:collfac_isr} and~\eqref{eq:collfac_fsr} than one would
have naively expected based on the known QCD formula. Nevertheless, it
turns out that there is a relation between the massive and massless
splitting functions. In particular, we find
\begin{subequations}\label{eq:massless_splittings}
\begin{align}
    &J^{(0)}(y,m) \wideeq{m\to 0} \bar{J}^{(0)}(y) + \mathcal{O}(m), \\
    &J^{(1)}(y,m) \wideeq{m\to 0}
    \bar{J}^{(1)}(y) + Z^{(1)}(m)\bar{J}^{(0)}(y)+ \mathcal{O}(m),
\end{align}
\end{subequations}
where the massive splitting function reduces in the massless limit to
the massless one plus singular corrections from massification. It is
conceivable that the same relation will also hold beyond one loop. In this case, however, there will be non-vanishing soft contributions from massification~\cite{Engel:2018fsb}. In addition to being an
interesting result in its own right, this represents a strong check
for the validity of the results presented in this section and in
Appendix~\ref{sec:splitfunc}.

\section{Validation}
\label{sec:validation}

To demonstrate the correctness and applicability of
equations~\eqref{eq:lbk_oneloop}, ~\eqref{eq:collfac_isr}
and~\eqref{eq:collfac_fsr} we consider the soft and collinear limits
in the process
\begin{align}\label{eq:dummyprocess}
e^-(p_1)e^+(p_2) \to e^-(p_3) e^+(p_4) \gamma(p_5)\gamma(k)
\end{align}
at one loop where $k$ can become soft or collinear. This $2\to4$
process is a highly non-trivial test of our formalism since the full
one-loop matrix element is rather involved and contains hexagon
functions. We will compare our approximations to
OpenLoops~\cite{Buccioni:2017yxi, Buccioni:2019sur} running in
quadruple precision mode~\cite{max}. This, while obviously slower, is
remarkably stable and produces reliable results deep into the soft and
collinear limits.

The process~\eqref{eq:dummyprocess} could also be considered to be the
real-real-virtual contribution to the N$^3$LO corrections to Bhabha
scattering. Hence, implementing this matrix element in a way that
remains sufficiently stable for collinear and soft emission would be
essential for any future N$^3$LO calculation.

In the following discussion we use a centre-of-mass energy of
$\sqrt{s}=10.583\,{\rm GeV}$, tailored to the beam energy of the Belle
II experiment.

\subsection{Soft limit}

\begin{figure}
    \centering
    \includegraphics[width=0.8\textwidth]{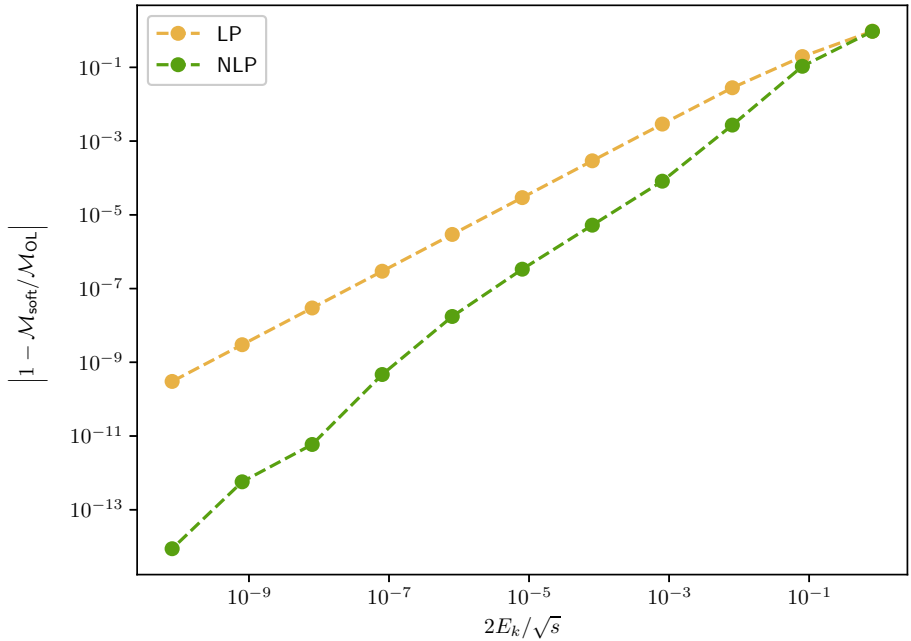}
    \caption{Convergence of the soft limit $\mathcal{M}_\text{soft}$ at leading and subleading
    power. The reference value $\mathcal{M}_\text{OL}$ is calculated with OpenLoops using
    quadruple precision.}
\label{fig:ee2eegg_nts}
\end{figure}

Let us begin by considering the limit where one of the two photons becomes
soft while the other photon remains hard, i.e. $k\to 0$.  Looking
at~\eqref{eq:lbk_oneloop}, we have
\begin{subequations}
\begin{alignat}{10}
&p_1&\to&+p_1,&\qquad   &p_2&\to&+p_2,&\qquad
&p_3&\to&-p_3,&\qquad   &p_4&\to&-p_4,&\qquad
&p_5&\to&-p_5,
\\
&Q_1&=  &- e,&   &Q_2&=  &+ e,&
&Q_3&=  &+ e,&   &Q_4&=  &- e,&
&Q_5&=   &\ 0,
\end{alignat}
\end{subequations}
Of course the above sign convention for the outgoing particles also
has to be taken into account in the case of the derivatives
$\partial/\partial p_{i,\mu}$. Furthermore, we define the set of
invariants $\{s\}$ as
\begin{align}
    \{s\}=\{s=(p_1+p_2)^2,
            t=(p_2-p_4)^2,
            s_{15}=2p_1 \cdot p_5, 
            s_{25}=2p_2 \cdot p_5,
            s_{35}=2p_3 \cdot p_5\}.
\end{align}
We emphasise again that this choice is not unique. It is therefore
crucial to use the same definition both in the evaluation of the
non-radiative matrix element as well as for the derivatives
$\partial/\partial p_{i,\mu}$ in the LBK operator~\eqref{eq:lbkop}.
Since already the one-loop matrix element for $ee\to ee\gamma$ is
rather complicated, we have implemented the corresponding derivatives
numerically to a very high precision in {\tt Mathematica}. Combining
this with the soft contribution from~\eqref{eq:lbk_oneloop_soft} then
yields the complete subleading power approximation. The corresponding
$\xi^{-2}$ and $\xi^{-1}$ terms can then be compared to OpenLoops as a
function of the `softness' $2E_k/\sqrt{s}$. The result is shown in
Figure~\ref{fig:ee2eegg_nts} down to values of $10^{-10}$. It is
clearly visible that including the $\xi^{-1}$ (NLP) terms
significantly improves the precision of the approximation. This
behaviour clearly validates our one-loop generalisation of the LBK
theorem presented in~\eqref{eq:lbk_oneloop}.

\subsection{Collinear limit}

\begin{figure}
    \centering
    \includegraphics[width=0.8\textwidth]{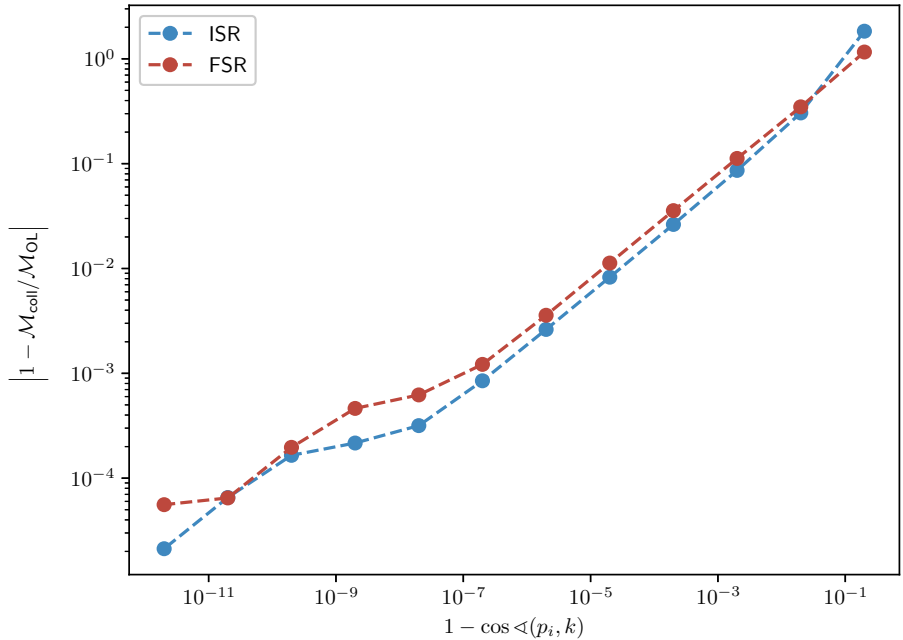}
    \caption{Convergence of the collinear limit
    $\mathcal{M}_{\text{coll}}$ at leading power for ISR
    \eqref{eq:rrv:isr} and FSR \eqref{eq:rrv:fsr} as a function of the
    collinearity. The reference value $\mathcal{M}_{\text{OL}}$ is
    calculated with OpenLoops in quadruple precision.}
\label{fig:ee2eegg_pcl}
\end{figure}

Next, let us consider the collinear limit. Once the massless
one-loop matrix element for the non-radiative process
$e^-(p_1)e^+(p_2) \to e^-(p_3) e^+(p_4) \gamma(p_5)$ is known, the
application of the factorisation formulas~\eqref{eq:collfac_isr}
and~\eqref{eq:collfac_fsr} is rather straightforward. As an example we
consider the case of the photon $k$ becoming collinear to $p_1$ (ISR)
or the case of it becoming collinear to $p_3$ (FSR). The cases of
$p_2$ and $p_4$ are completely analogous.

The massified approximation~\cite{Engel:2018fsb} for the matrix
element reads
\begin{align}
\mathcal{M}_{n+1} \wideeq{m\sim\lambda}
    Z(m_1) Z(m_2) Z(m_3) Z(m_4)
    \mathcal{M}_{n+1}(p_1, p_2, p_3, p_4, p_5, k; m_i=0)
    +\mathcal{O}(\lambda) 
\end{align}
which is valid for the bulk of the phase space, i.e. assuming $k$ is
neither soft nor collinear. Note that the masses are only given
indices so that the different $Z$ can be better disentangled once $k$
becomes collinear. Of course all $m_i$ are equal.

In the ISR limit we replace $Z(m_1)$ with $J_\text{ISR}$,
reducing the number of particles in the massless matrix element
\begin{align}
\mathcal{M}_{n+1} \wideeq{k\cdot p_1\sim\lambda^2}
    \frac{J_\text{ISR}(x, m_1)}{\lambda^2} Z(m_2) Z(m_3) Z(m_4)
    \mathcal{M}_{n}(p_1-k, p_2, p_3, p_4, p_5; m_i=0)
    +\mathcal{O}(\lambda^{-1}).\label{eq:rrv:isr}
\end{align}
In complete analogy the FSR limit is given by
\begin{align}
\mathcal{M}_{n+1} \wideeq{k\cdot p_3\sim\lambda^2}
    Z(m_1) Z(m_2) \frac{J_\text{FSR}(z, m_3)}{\lambda^2} Z(m_4)
    \mathcal{M}_{n}(p_1, p_2, p_3+k, p_4, p_5; m_i=0)
    +\mathcal{O}(\lambda^{-1}).\label{eq:rrv:fsr}
\end{align}
All that is left to do before we can compare to OpenLoops is to
multiply out the terms in \eqref{eq:rrv:isr} and \eqref{eq:rrv:fsr}.
The result of this comparison is shown in Figure~\ref{fig:ee2eegg_pcl}
for ISR and FSR as a function of the `collinearity'
$1-\cos\sphericalangle(p_i,k)$. To understand the observed convergence
behaviour it is important to realise that the expansion is not performed
in the collinearity but in $k\cdot p_i, m_i^2 \to 0$. The approximation
thus only improves while $k\cdot p_i$ gets smaller. At the point,
however, where $k\cdot p_i$ approximately satisfies~\eqref{eq:kp_scaling}, the limit saturates since $m_i$
is kept constant. This explains the kink at $10^{-7}$. We can therefore
conclude that Figure~\ref{fig:ee2eegg_pcl} represents a strong
validation of our factorisation formulas given in~\eqref{eq:collfac_isr}
and~\eqref{eq:collfac_fsr}.

\section{Conclusion and outlook}
\label{sec:conclusion}

In this paper we have presented two novel findings about the universal
structure of radiative QED amplitudes in the soft and in the collinear
limit. We have extended the well-known LBK theorem to
the one-loop level, relating the radiative with the non-radiative
amplitude at subleading power in the soft limit. The additional loop
effects are taken into account by supplementing the LBK theorem with a
soft function that we have evaluated in a universal way. In addition, we
have derived a factorisation formula at one loop that
describes the leading-power collinear limit in the presence of small
but non-vanishing fermion masses. Contrary to the analogous result for
massless QCD we also get contributions from non-collinear light
external fermions. These additional terms in the factorisation formula
take into account the corresponding leading small-mass effects.

The approximation of real-virtual amplitudes with the soft limit at
subleading power can be used to achieve a stable and efficient
implementation of this contribution. In the case of Bhabha and
M{\o}ller scattering this next-to-soft stabilisation enabled the
first fully differential NNLO calculation.  Even though the
computation of the soft expansion was straightforward it turned out to
be cumbersome. The extension of the LBK theorem presented in this
paper significantly facilitates the application of the next-to-soft
stabilisation method to other processes.

This is particularly relevant in light of the MUonE
experiment~\cite{CarloniCalame:2015obs,Abbiendi:2016xup,Abbiendi:2677471} where a theory prediction at the
level of 10ppm is needed to achieve the targeted experimental
precision. The minimal requirement for this is a fixed-order NNLO QED
Monte Carlo matched to a parton shower. This has triggered a wide
theory effort~\cite{Banerjee:2020tdt} where many partial results have
been calculated in the past years. Two Monte Carlo codes are available
that include the dominant electron-line corrections at NNLO~\cite{CarloniCalame:2020yoz,Banerjee:2020rww}. In addition, the
subset of the NNLO corrections with closed (and open) fermion loops
are also known~\cite{Fael:2018dmz,Fael:2019nsf,Budassi:2021twh}.
Further, electroweak effects~\cite{Alacevich:2018vez} and possible
contaminations from physics beyond the Standard
Model~\cite{Schubert:2019nwm,Dev:2020drf,Masiero:2020vxk} have been studied. Very 
recently a
crucial step towards the full set of NNLO QED corrections was
accomplished with the calculation of the two-loop amplitude with a
non-zero muon mass~\cite{Bonciani:2021okt}.
After massification~\cite{Engel:2018fsb} of the massless electron this result can be included in a Monte Carlo code. The remaining
bottleneck is therefore a numerically stable implementation of the
real-virtual contribution. Next-to-soft stabilisation combined with
the extension of the LBK theorem presented in this paper represents an
elegant solution to this problem.

One could in principle also follow a similar approach in the case of
the collinear limit. Contrary to the soft limit, however, the
reliability of the approximation is limited by the scale hierarchy
between the light fermion mass and the typical energy scale of the
considered process. Leading-collinear stabilisation could therefore
only be used reliably in the case of high energies. On the other hand,
also for low-energy processes the small electron mass leads to large
peaks in the radiative amplitudes in collinear regions complicating
the numerical integration over the phase space. The collinear
factorisation formula presented in this paper could thus be used as
basis for a subtraction scheme for these collinear
pseudo-singularities similar to the NLO formalism developed
in~\cite{Dittmaier:1999mb}.

In principle, the same strategies applied in this paper can also be used to extend the next-to-soft and leading-collinear factorisation formulas beyond one loop. However, a more formal understanding of the presented results in terms of an effective field theory could facilitate this task. For the LBK theorem~\eqref{eq:lbk_oneloop} this would entail a definition of the soft function~\eqref{eq:lbk_oneloop_soft} in the framework of HQET. In the case of the collinear splitting function, on the other hand, a construction in the context of SCET would be needed.

Finally, we remark that this paper makes it possible to calculate
contributions from radiative amplitudes based on the corresponding
massless result. In order to do so one would use
massification~\cite{Engel:2018fsb} for the bulk of the phase space and
otherwise switch to the next-to-soft or leading-collinear
approximations. While this is not strictly necessary at one loop it
will be indispensable at two loop. The extension of the presented
results for radiative two-loop amplitudes is therefore planned for the
future. This in turn would represent a major step towards the
fully differential calculation of the dominant N$^3$LO corrections for
muon-electron scattering.

\subsection*{Acknowledgement} 

We are very grateful to M. Zoller for providing the real-real-virtual
matrix element for Bhabha scattering in OpenLoops and for assisting
with its proper usage. Special thanks go to N. Schalch whose master
thesis has triggered the studies presented in this paper. Furthermore,
we would like to thank T. Becher and A. Broggio for useful
discussions. TE acknowledges support by the Swiss National Science
Foundation (SNF) under contract 200021\_178967. YU acknowledges
support by the UK Science and Technology Facilities Council (STFC)
under grant ST/T001011/1.

\newpage
\begin{appendix}
\label{sec:appendix}

\section{Soft integrals}\label{sec:softints}

In~\eqref{eq:softinta} and~\eqref{eq:softintb} we have defined the two integrals necessary to
construct the soft contribution. These integrals
are universal and are given here in $d=4-2\epsilon$ dimensions with $\mu$ denoting the scale of dimensional regularisation. With
$s_{ij} = 2p_i\cdot p_j$, $s_{i\gamma} = 2 k\cdot p_i$, and
\begin{align}
    C(\epsilon) = \frac{(4\pi)^\epsilon}{16\pi^2 \Gamma(1-\epsilon)}
\end{align}
we have
\begin{subequations}
\begin{align}
I_1(p_i,k) &= i\int \frac{\text{d}^d \ell}{(2\pi)^d} 
        \frac{1}{[\ell^2+i\delta][\ell\cdot p_i-k\cdot p_i+i\delta]} \\
 &= -2C(\epsilon)\bigg(\frac{m_i^2}{s_{i\gamma}}+i\delta\bigg)^{2\epsilon-1}
    \bigg(\frac{\mu^2}{m_i^2}\bigg)^\epsilon
    \Gamma(1-\epsilon)^2\Gamma(2\epsilon-1),
\\
I_2(p_i,p_j,k) &
    = i\int \frac{\text{d}^d \ell}{(2\pi)^d} 
        \frac{1}{[\ell^2+i\delta][-\ell\cdot p_j+i\delta][\ell\cdot p_i-k\cdot p_i+i\delta]}
    \\&= 
  \frac{8C(\epsilon)}{s_{ij}}
  \bigg(\frac{m_i^2}{s_{i\gamma}}+i\delta\bigg)^{2\epsilon}
  \bigg(\frac{\mu^2}{m_i^2}\bigg)^\epsilon
  \bigg(\frac{|s_{ij}|}{2m_im_j}\bigg)^{2\epsilon}
  \Bigg\{
    \Gamma(1-\epsilon)^2\Gamma(2\epsilon-1)
    \ \pFq{2}{1}{\tfrac12-\epsilon,1-\epsilon}{\tfrac32-\epsilon}{v_{ij}^2}
    \notag\\&\qquad\qquad\qquad
   + i\pi
    \Big(-\frac14+i\delta\Big)^{-\epsilon}
    v_{ij}^{-1+2\epsilon}
    \Gamma(1-2\epsilon)\Gamma(2\epsilon)
    \Theta(s_{ij})
  \Bigg\}\\
  &= \frac{4C(\epsilon)}{s_{ij}}\frac{1+\chi}{1-\chi}
  \bigg(\frac{m_i^2}{s_{i\gamma}}+i\delta\bigg)^{2\epsilon}
  \bigg(\frac{\mu^2}{m_i^2}\bigg)^\epsilon
  \Bigg\{
    \frac{H_0(\chi)}{2\epsilon}-\zeta_2-\frac{1}{2}H_{0,0}(\chi)-H_{1,0}(\chi)
    \notag\\&\qquad\qquad\qquad
   + \Theta(s_{ij}) 6\zeta_2
    +i \pi \Theta(s_{ij}) \Big(\frac{1}{\epsilon}-H_{0}(\chi)-2H_{1}(\chi) \Big)
    +\mathcal{O}(\epsilon)
  \Bigg\},
\end{align}
\end{subequations}
where $v_{ij}=\sqrt{1-4m_i^2m_j^2/s_{ij}^2}=(1-\chi)/(1+\chi)$. The {\tt Mathematica} package {\tt
HypExp}~\cite{Huber:2007dx,Huber:2005yg} was used to expand the hypergeometric
function in $I_2$ in terms of the harmonic polylogarithms (HPLs) of
Remiddi and Vermaseren~\cite{Remiddi:1999ew}. We then applied the
related {\tt Mathematica} code {\tt
HPL}~\cite{Maitre:2005uu,Maitre:2007kp} to simplify the resulting
expression. Since $0<v_{ij}<1$ and thus $0<\chi<1$ all HPLs are
manifestly real.

\section{Splitting functions}\label{sec:splitfunc}

In the following we give the explicit expressions for all the
quantities that enter the collinear factorisation
formulas~\eqref{eq:collfac_isr} (ISR) and~\eqref{eq:collfac_fsr}
(FSR). The results are presented in a form that can be used in three
major flavours of dimensional regularisation: the four-dimensional
helicity scheme (\FDH), 't Hooft-Veltman scheme (\HV), and
conventional dimensional regularisation (\CDR)
(see~\cite{Gnendiger:2017pys} and references therein for the
definitions of these schemes). To this end, we keep the dimensionality
of $\epsilon$ scalars, $n_\epsilon$, explicit in the poles but set it to
zero in the finite parts. The regularisation-scheme dependence is
therefore manifest as terms $\propto n_\epsilon$.  The corresponding
results in \HV and \CDR can be obtained by setting $n_\epsilon=0$.
Inserting $n_\epsilon=2\epsilon$, on the other hand, retrieves the
expressions in \FDH. Furthermore, in the case of \HV and \FDH
$\epsilon$ has to be set to zero in the tree-level splitting function.

In agreement with~\cite{Engel:2018fsb} the massification
factor reads
\begin{align}
    Z^{(1)} =
    4\pi\alpha C(\epsilon)
    \Big(\frac{\mu^2}{m^2}\Big)^\epsilon
    \Big(
    \frac{2}{\epsilon^2}
    +\frac{1}{\epsilon}(1-\frac{1}{2}n_\epsilon)
    +4+2\zeta_2
    \Big)
    +\mathcal{O}(\epsilon),
\end{align}
with $\alpha = Q^2/(4\pi) = e^2/(4\pi)$.

We then define the invariant $s_{kp}=2k\cdot p$ where the photon
momentum $k$ is collinear to an initial- or final-state fermion $p$.
The initial-state collinear splitting function
\begin{align}
    J_\text{ISR} 
    = J_\text{ISR}^{(0)}+J_\text{ISR}^{(1)}+\mathcal{O}(\alpha^2)
\end{align}
can conveniently be written in terms of
\begin{align}
    x=\frac{E_p-E_k}{E_p},\qquad u=\frac{s_{kp}-m^2}{s_{kp}},
\end{align}
as
\begin{align}
J_\text{ISR}^{(0)}
=& \frac{8\pi\alpha(1-u)}{m^2(1-x)x}
\big(1-2x+3x^2+2xu-2x^2u-\epsilon(1-x)^2\big), \\
J_\text{ISR}^{(1)}
=& 8\pi\alpha J_\text{ISR}^{(0)} C(\epsilon)\Big(\frac{\mu^2}{m^2}\Big)^\epsilon
\Big(
\frac{1}{\epsilon^2} 
+\frac{1}{4\epsilon} (2-8 H_{0}(x)-n_\epsilon)
\Big)
+ \alpha^2 \frac{4(1-u)}{x m^2} \tilde{J}_\text{ISR}
+\mathcal{O}(\epsilon),
\end{align}
with
\begin{align}
\begin{split}
\tilde{J}_\text{ISR}
=& \frac{1}{u(1-x)}(-2 u^2 x^2+2 u^2+2 u x^2+2 u x+x^2-x) \\
+& \frac{\zeta_2}{1-x}(2 u^2 x^2-4 u^2 x+2 u^2-12 u x^2+16 u x-4 u+13 x^2-12 x+5) \\
+& \frac{H_{1}(u)}{u^2}(2 u^3 x-5 u^2 x-2 u^3+3 u^2+2 u x-u+x) \\
+& \frac{H_{1}(u)H_{0}(x)+H_{1,0}(x)}{1-x}(-8 u x^2+8 u x+10 x^2-8 x+2) \\
+& \big(H_{2}(u)+H_{1,1}(u)\big)(-2 u^2 x+2 u^2+2 u x-4 u+2).
\end{split}
\end{align}
Because $0<x<1$ and $u<1$ the above expression is always real. 

The result for the final-state splitting function
\begin{align}
    J_\text{FSR} 
    = J_\text{FSR}^{(0)}+J_\text{FSR}^{(1)}+\mathcal{O}(\alpha^3)
\end{align}
can be obtained from $J_\text{ISR}$ via the crossing relation
$p\to-p$. In particular, this implies $x \to z^{-1}$ and $u \to
v^{-1}$ with
\begin{align}
    z=\frac{E_p}{E_p+E_k},\qquad v=\frac{s_{kp}}{m^2+s_{kp}}.
\end{align}
The corresponding analytic continuation is unambiguously defined via
the prescription $s_{kp} \to s_{kp} + i \delta$ or equivalently $u \to
u - i \delta$. We then find
\begin{align}
J_\text{ISR}^{(0)}
=& \frac{8\pi\alpha(1-v)}{m^2v^2(1-z)}\big(v z^2-2 v z+3 v+2 z-2-\epsilon v(1-z)^2\big), \\
J_\text{FSR}^{(1)}
=& 8\pi\alpha J_\text{FSR}^{(0)} C(\epsilon)\Big(\frac{\mu^2}{m^2}\Big)^\epsilon
\Big(
\frac{1}{\epsilon^2} 
+\frac{1}{4\epsilon} (2+8 H_{0}(z)-n_\epsilon)
\Big)
+ \alpha^2 \frac{4(1-v)}{m^2v^2} \Big( \tilde{J}_\text{FSR}^\text{Re}+i\pi \tilde{J}_\text{FSR}^\text{Im} \ \Big)
+\mathcal{O}(\epsilon),
\end{align}
with
\begin{align}
\begin{split}
\tilde{J}_\text{FSR}^\text{Re}
=& \frac{1}{1-z}(-v^2z+v^2+2 v z+2 v+2 z^2-2) \\
+& \frac{\zeta_2}{1-z}(-v z^2+6 v z-7 v-6 z+6) \\
+&\big(H_{0}(v)+H_{1}(v)\big)(v^2 z-v^3-2 v^2-3 v z+5 v+2 z-2) \\
+& \frac{H_{0}(v)H_{0}(z)+H_{1}(v)H_{0}(z)+H_{0,0}(z)+H_{1,0}(z)}{1-z}
(-2 v z^2+8 v z-10 v-8 z+8) \\
+& \frac{H_{1,0}(v)+H_{1,1}(v)}{v}(-2 v^2 z+4 v z-2 v-2 z+2),
\\ \vspace{.1cm}
\tilde{J}_\text{FSR}^\text{Im}
=& -v^2 z+v^3+2 v^2+3 v z-5 v-2 z+2 \\
+& \frac{H_{0}(z)}{1-z}(2 v z^2-8 v z+10 v+8 z-8) \\
+& \frac{H_{1}(v)}{v}(2 v^2 z-4 v z+2 v+2 z-2).
\end{split}
\end{align}
The imaginary part is given explicitly leaving all of the HPLs real for the physical region where $0<z,u<1$. The massless version of the FSR splitting function entering~\eqref{eq:massless_splittings} can be extracted from the spin-summed result of equations (II.10) and (II.11) in~\cite{Bern:1994zx} by taking the QED limit. The corresponding expressions in the \FDH scheme read 
\begin{align}
    &\bar{J}_\text{FSR}^{(0)} =
    \frac{8\pi\alpha}{s_{kp}} \frac{1+z^2}{1-z}, \\
    & \bar{J}_\text{FSR}^{(1)} =
    16\pi\alpha C(\epsilon) \Big\{ \bar{J}_\text{FSR}^{(0)} 
    \Big(
    -\frac{1}{\epsilon^2}\Big(-\frac{\mu^2}{z s_{kp}}\Big)^\epsilon
    +\frac{1}{\epsilon^2}\Big(-\frac{\mu^2}{s_{kp}}\Big)^\epsilon
    -H_{1,0}(z)-\zeta_2
    \Big)
    -\frac{4\pi\alpha}{s_{kp}} \Big\}.
\end{align}

\end{appendix}

\bibliographystyle{JHEP}
\bibliography{limits.bib}

\end{document}